\documentclass[12pt]{JHEP3}
\usepackage{amsmath,amssymb}
\usepackage{epsfig}

\def\x'{\mathaccent 19 x}
\def\y'{\mathaccent 19 y}
\def\n'{\mathaccent 19 n}
\def\u'{\mathaccent 19 u}

\def\et'{\mathaccent 19 \eta}
\def\th'{\mathaccent 19 \theta}
\def\lam'{\mathaccent 19 \lambda}
\def\varet'{\mathaccent 19 \vartheta}
\def\rh'{\mathaccent 19 \rho}
\def\ph'{\mathaccent 19 \Phi}
\def\xb'{\mathaccent 19 {\bar{x}}}

\def\be{\begin{equation}}
\def\ee{\end{equation}}

\newcommand{\bea}{\begin{eqnarray}}
\newcommand{\eea}{\end{eqnarray}}

\def\Tr{\text{Tr}}

\author{Antonio Bassetto$^{(a)}$, Luca Griguolo$^{(b)}$, Fabrizio Pucci$^{(c)}$ and Domenico~Seminara$^{(c)}$\\
$^{(a)}$ Dipartimento di  Fisica, Universit\`a  di Padova and
INFN Sezione di Padova,\\ Via Marzolo 8, 31131 Padova, Italy \\
$^{(b)}$  Dipartimento di  Fisica, Universit\`a  di Parma and
INFN Gruppo Collegato di Parma, Viale G.P. Usberti 7/A, 43100 Parma, Italy\\
$^{(c)}$ Dipartimento di Fisica, Universit\`a di
Firenze and INFN Sezione di Firenze, Via  G. Sansone 1, 50019 Sesto Fiorentino, Italy\\

\email{bassetto@pd.infn.it, griguolo@fis.unipr.it, pucci@fi.infn.it, seminara@fi.infn.it}}

\abstract{We study the quantum properties of certain BPS
Wilson loops in ${\cal N}=4$ supersymmetric Yang-Mills theory. They belong to a
general family, introduced recently, in which the addition
of particular scalar couplings endows generic loops on $S^3$ with a fraction of supersymmetry.
When restricted to $S^2$, their quantum average has been further conjectured to be exactly computed by the matrix model
governing the zero-instanton sector of YM$_2$ on the sphere. We perform a complete two-loop analysis on a class
of cusped Wilson loops lying on a two-dimensional sphere, finding perfect agreement with the conjecture.
The perturbative computation reproduces the matrix-model expectation through a highly non-trivial interplay
between ladder diagrams and self-energies/vertex contributions, suggesting the existence of a localization procedure.   }

\title{Supersymmetric Wilson loops at two loops}
\preprint{}
\begin{document}

\renewcommand{\thefootnote}{\arabic{footnote}}
\setcounter{footnote}{0}

\section{Introduction}

The $AdS/$CFT correspondence \cite{Maldacena:1997re,Gubser:1998bc,Witten:1998qj} is a particularly striking example of relation between gauge theories
in four dimensions and string theories: ${\cal N}=4$ supersymmetric Yang-Mills (SYM) theory is expected be dual to type IIB superstring theory
on $AdS_5\times S^5$. In order to check this powerful connection one would like to compare results for
the same observables as obtained from the different sides of the correspondence. Unfortunately,
while it is relatively easy to compute quantities at weak coupling through familiar gauge techniques
and at strong coupling by exploiting string methods, there is no overlap between the regions of validity of
the two calculations. The original checks of the conjecture have been therefore restricted to compare
highly protected quantities, such as correlation functions of chiral operators, where the full complexity
of perturbation theory does not need to be taken into account.

In the last years the situation has experienced a dramatic improvement with the discovery of integrability in
${\cal N}=4$ SYM at large $N$ \cite{Minahan:2002ve}: Bethe ansatz techniques applied to the computation of anomalous dimensions
of local operators have opened the possibility to extrapolate results from weak to strong coupling.
The most astonishing example involves the so-called cusp anomaly, whose nonperturbative expression is encoded into
an exact integral equation \cite{Beisert:2006ez}: the weak coupling perturbative solution agrees with the Feynman diagram expansion
\cite{Bern:2006ew} while the strong coupling asymptotic solution \cite{Benna:2006nd}-\cite{Kostov:2008ax} reproduces the sigma-model result obtained from string theory \cite{Gubser:2002tv}-\cite{Roiban:2007jf}.

More recently we have seen also an impressive advance in studying scattering amplitudes in ${\cal N}=4$ SYM at
large $N$: a particulary intriguing conjecture for the all-order form of MHV $n$-gluon amplitudes has been proposed
in \cite{Bern:2005iz}, starting from the weak coupling expansion. The BDS conjecture was found to be consistent with the string
computation performed at strong coupling by Alday and Maldacena \cite{Alday:2007hr} for $n=4$. Quite surprisingly Wilson loops
play a central role in this last development, the amplitude itself being calculated from light-like loops in string
theory. This unexpected relation, appearing in string theory from a T-dual description of the scattering process,
holds even at weak coupling \cite{Drummond:2007au,Drummond:2007cf,Drummond:2007aua,Brandhuber:2007yx} and survives the recent six-gluons calculations \cite{Drummond:2007bm,Bern:2008ap,Drummond:2008aq} while disproving the BDS conjecture in its original form. The cusp anomaly is also part of this story, appearing directly in the amplitudes through divergent and finite terms: actually its original definition was exactly given in terms of light-like Wilson loops \cite{Korchemsky:1987wg,Korchemsky:1992xv}.

The importance of Wilson loops in checking $AdS$/CFT correspondence emerges also in a different computation,
that represents an older example of interpolation between weak and strong coupling: the circular Wilson loop, whose
exact expectation value, calculated from the gauge theory side, appears to be captured by a matrix model \cite{Erickson:2000af,Drukker:2000rr}, encoding the full perturbative expansion. This result is consistent with the $AdS$/CFT prediction, the strong coupling limit being precisely reproduced by string computations including an infinite series of $1/N$ corrections \cite{Drukker:2005kx,Drukker:2006zk,Yamaguchi:2006tq} (see also \cite{Gomis:2006sb}-\cite{Giombi:2006de}). Nevertheless the belief that a matrix model controls the circular Wilson loops was largely based on the original two-loop computation \cite{Erickson:2000af}, showing that in Feynman gauge only exchange diagrams contribute to the quantum average, self-energies and vertex diagrams describing the true SYM interactions summing to zero in the final result. A subsequent argument \cite{Drukker:2000rr}, based on the conformal relation between the circle and the trivial straight line, was often advocated to justify the assumption that interacting diagrams decouple at all orders, at least in Feynamn gauge, but a clear reason for this astonishing cancelation was missing. Doubts on this all-order behavior, based on Wilson loops correlators at three-loops, were raised in \cite{Plefka:2001bu,Arutyunov:2001hs} while it was soon realized that there are many matrix models, with the same two-loop expansion, leading to the string result at strong coupling \cite{Akemann:2001st}.

On the other hand the circular Wilson loop is invariant under a particular subset of superconformal transformations, generated by
combinations of $Q$'s and $S$'s charges: with respect to the full superconformal group it is a 1/2 BPS object. For long time it was suspected
that the deep reason behind the exact matrix model computation should be found in the BPS property: in particular the vanishing of interacting
diagrams suggested the existence of some twisted version of the theory \cite{Vafa:1994tf}, making the circular Wilson loop a topological observable.
Recently it was proved \cite{Pestun:2007rz} that this is indeed the case: the theory can be formulated on $S^4$ in such a way that the path integral localizes on a finite dimensional space and reduces to a simple gaussian matrix model. The circular Wilson loop, due to its invariance properties, can be computed as an observable in this matrix model, leading to the expected result: quite interestingly instanton corrections are claimed
to be absent. The author has studied also the more general situation of ${\cal N}=2,2^*$ theories obtaining localization on more complicated matrix models and instanton corrections.

It is very tempting therefore to study generalizations of the circular Wilson loops, carrying some amount of superconformal
invariance: they could generate new exact computations at gauge theory level, in principle testable at strong coupling by string theory.
An important step in this direction has been taken in \cite{Drukker:2007qr}: the authors have been able to construct a family of supersymmetric Wilson loop operators in ${\cal N}=4$ SYM, modifying the scalar couplings with the geometrical contour. For a generic curve on an $S^3$ in space-time
the loop preserves two supercharges but they discussed special cases which preserves 4, 8 and 16 supercharges. They also found for certain loops
explicit dual string solutions. Of particular interest are the loops restricted to $S^2$ because a one-loop computation suggests the
equivalence with analogous observables in purely bosonic Yang-Mills theory on the sphere. Two-dimensional Yang-Mills theory on compact surfaces can also be solved by localization \cite{Witten:1992xu} and one would suspect the methods presented in \cite{Pestun:2007rz} for the circular loop may be extended to the present situation. Another interesting feature is that the contours are not constrained to be smooth and possible links with the cusp anomaly may be explored.

To pursue the above program it is important to check the equivalence between supersymmetric Wilson loops on $S^2$ and two-dimensional Yang-Mills observables beyond the leading perturbative order: even at two loops this involves a non-trivial technical computation. Basically one should calculate the fourth-order contribution, in closed form, to a generic contour lying on a sphere embedded in space-time: all the loops in
this family are, in particular, non-planar. The strategy followed in the case of the circle \cite{Erickson:2000af} was first to prove finiteness of
a generic loop, with the usual scalar coupling, in the Feynman gauge and then to show, in the circular case, the cancelation between self-energy diagrams and vertex diagram, without computing any integrals at this order. The sum of ladder diagrams can be instead performed rather
easily, being reduced to a matrix model problem because the effective propagators turns out to be constant. The situation here appears far more complicated: it is not guaranteed to find similar properties and the conjectured result could be recovered through a delicate combination of ladder and interacting diagrams.

In this paper we are mainly concerned with this problem and we perform the two-loop computation needed to support the conjecture. We will restrict ourselves to a class of contours made by two longitudes separated by an arbitrary angle: these contours have cusps but, as we will see later, no divergence arises thanks to the peculiar scalar couplings. We find that the equivalence with Yang-Mills theory on the sphere survives at two loops in a rather non-trivial way: in fact
only when ladder diagrams are combined with self-energy and vertex graphs the expected result is recovered.
The plan of the paper is the following:
in section 2 we briefly review the properties of the family of loops introduced in \cite{Drukker:2007qr}, paying particular attention to tha case of $S^2$, and we explain the details of the conjectured relation with YM$_2$. Section 3 is devoted to the perturbative expansion of supersymmetric
Wilson loops: in particular we derive a compact formula that encodes the contribution of self-energy and vertex diagrams to the two-loop expansion of  supersymmetric Wilson loops in ${\cal N}=4$ SYM. It gives a manifestly finite result and it is based on a particularly intriguing ``subtraction"
procedure, suggested by the light-cone gauge formulation of the theory. In section 4 we specialize our computations to the case of the cusped loops on $S^2$: by a mixed combination of analytical and numerical calculations we obtain the fourth-order contribution, finding exact agreement with the two-dimensional theory. In section 5 we report on our conclusions and possible future developments. The relation of the formulas presented in section 3 with the light-cone gauge is the subject of Appendix A. Appendix B contains instead some technical material on the relevant integrals.
After submission, another paper  \cite{Young:2008ed} appeared  addressing the same issues.

\section{Supersymmetric Wilson loops and YM$_2$}
In the context of $AdS$/CFT correspondence it is quite natural to consider generalizations of the
familiar Wilson loop operator: the gauge multiplet of ${\cal N}=4$ SYM includes besides one gauge field,
six real scalars and four complex spinor fields and it is then possible
to incorporate them through suitable couplings to the contour. We
will consider here the extra coupling of the scalars
$\Phi^I$ (with $I=1,\cdots,6$) leading to the following expression \cite{Rey:1998ik,Maldacena:1998im}
for the Wilson loop
\begin{equation}
W=\frac{1}{N}\,\Tr\,{\cal P}\exp \oint_C dt
(iA_\mu\dot x^\mu(t) + |\dot x|\Theta^I(t)\Phi^I)\,,
\label{Wilson-loop}
\end{equation}
where $x^\mu(t)$ is the path of the loop and $\Theta^I(t)$ are arbitrary
couplings. While this choice is reminiscent of the ten
dimensional origin of the observable, we remark that fermionic couplings can be also considered
\cite{Bianchi:2002gz}.

To generically preserve some SUSY one should require that the norm of $\Theta^I$ be
one, but that alone is only locally sufficient. If one considers
the supersymmetry variation of the loop, then at every point along the loop
one finds different conditions for preserving supersymmetry. Only if all those
conditions commute, will the loop be globally supersymmetric. This happens, for example,
in the case of the straight line, where $\dot x^\mu$ is a
constant vector and one takes also $\Theta^I$ to be constant: at every point one finds indeed
the same equation. This possibility has been generalized by Zarembo \cite{Zarembo:2002an},
who assigned for every tangent vector in ${\mathbb R}^4$ a unit vector in ${\mathbb R}^6$ by a $6\times 4$ matrix $M^I{}_\mu$
and took $|\dot x|\Theta^I=M^I{}_\mu\dot x^\mu$. If the contour is contained within a one-dimensional linear
subspace of ${\mathbb R}^4$, half of the super-Poincar\'e symmetries
generated by $Q$ and $\bar Q$ are preserved, while inside a 2-plane $1/4$ of them survives. More
generally inside ${\mathbb R}^3$ we have $1/8$ SUSY and for a generic curve $1/16$.

An interesting property of those loops is that their expectation values seem to be trivial,
with evidence both from perturbation theory, from $AdS$/CFT duality and from
a topological argument \cite{Zarembo:2002an,Guralnik:2003di,
Guralnik:2004yc,Dymarsky:2006ve,Kapustin:2006pk}. Although surprisingly this
makes these observables somehow trivial if one would explore the interpolation
between weak and strong coupling in ${\cal N}=4$. On the other hand it is clear that some amount
of supersymmetry should be present in order to get exact results: one can therefore resort to different
loops, preserving some combination of the super-Poincar\'e and the super-conformal symmetries, generated by $S$ and $\bar S$.

The first celebrated example is the circular Wilson loop \cite{Erickson:2000af,Drukker:2000rr}: the contour $x^\mu(t)=(R \cos t,R \sin t,0,0)$ parameterizes a (euclidean) circle while $\Theta^I$ is a constant unit vector in ${\mathbb R}^6$. Its BPS properties are simply understood: the vacuum of ${\cal N}=4$ SYM has 32 supercharges generated by the spinors
\begin{equation}
\label{conformal spinor}
\epsilon(x)=\epsilon_0 +
x^\mu\gamma_\mu\epsilon_1\,,
\end{equation}
where $\epsilon_0$ is related to the Poincar\'e supersymmetries and
$\epsilon_1$ is related to the super-conformal ones. Here $\gamma_\mu$ are the usual Dirac matrices of $SO(4)$
while we will denote by $\rho^I$ the $SO(6)$ Dirac matrices, acting on the $R$-symmetry index of $\epsilon$
(the two sets of matrices are taken to mutually anti-commute). At the linear order the supersymmetry variation of the
Wilson loop is proportional to
\begin{equation}
(-i\gamma_1 \sin t+i \gamma _2 \cos t+\rho_3
)\epsilon(x).
\end{equation}
The above combination vanishes when
\begin{equation}
\epsilon_0=R\rho_3\gamma_1\gamma_2\epsilon_1:
\end{equation}
these configurations preserve 1/2 of the supersymmetries, always involving super-conformal
transformations. The quantum behavior of this class of loops is far more interesting than
their straight line cousins. For the circular loop, conformal invariance predicts that the
expectation value of the loop operator is independent of the radius of
the circle. In their seminal paper \cite{Erickson:2000af} Erickson, Semenoff and Zarembo  found
that the sum of all planar Feynman diagrams which have no internal vertices (which includes both rainbow
and ladder diagrams) in the 't Hooft limit produces the expression
\begin{equation}
\left< W_C\right>_{\rm ladders}=
\frac{2}{\sqrt{g^2N}}\,I_1(\sqrt{g^2N}),
\label{circladder1}
\end{equation}
where $I_1$ is the Bessel function.  Taking the large $g^2N$ limit
gives
\begin{equation}
\langle W_C\rangle_{\text{ladders}}
=
\frac{\mathrm{e}^{\sqrt{g^2N}}}{(\pi/2)^{1/2}(g^2N)^{3/4}},
\label{circpert1}
\end{equation}
which has an exponential behavior identical to the prediction of the
$AdS$/CFT correspondence \cite{Dru99,Ber98},
\begin{equation}
\left< W_C\right>_{\rm AdS/CFT}
=
\mathrm{e}^{\sqrt{g^2N}}.
\label{circpert2}
\end{equation}

It is intriguing that this sum of a special class of diagrams produced
the exact asymptotic behavior that is predicted by the $AdS$/CFT
correspondence, considering that it does not include any diagrams
which have internal vertices.  Assuming that the $AdS$/CFT prediction is
indeed the correct asymptotic behavior, one could optimistically guess that
corrections to the sum of ladder diagrams cancel and the result
(\ref{circladder1}) is exact. This expectation was checked in \cite{Erickson:2000af}
by computing the leading order corrections to (\ref{circladder1}) coming
from diagrams with internal vertices.  These occur at order $g^4N^2$ and
these diagrams do indeed cancel exactly when the spacetime dimension is
four. Assuming the vanishing of the interacting diagrams at higher-order
at finite $N$ either (consistently with the fourth-order calculations ) an exact expression
for the Wilson loop, by summing the ladder diagrams, has been further proposed \cite{Drukker:2000rr}
\begin{equation}
<{W_C}>=
<{{1\over N}\Tr\exp(M)}>
={1\over Z}\int {\cal D} M {1\over N}\Tr\Bigl[\exp(M)\Bigr]
\exp\left(-{2\over g^2}\Tr M^2\right)\,.
\label{matrix}
\end{equation}
We see that the exact value of the Wilson loop is obtained by computing a particular
observable in a gaussian matrix model, encoding the full resummation of the
perturbative series. We stress that this expression has been argued on the basis of
some properties of the theory in Feynman gauge: the basic assumption that interacting
diagrams give vanishing contribution has been checked indeed in this gauge. Moreover
the ladder diagrams can be summed by the matrix model (\ref{matrix}) because of the peculiar
structure of the combined vector-scalar propagator: let us consider the $2n$-th order term in the
Taylor expansion of the loop
\[
\frac{1}{N}
\int_0^{2\pi} dt_1\int_0^{t_1}dt_2\cdots
\int_0^{t_{2n-1}}dt_{2n}  \Tr
\bigl<\bigl(iA(t_1)+\Phi(t_1)\bigr)
\cdots
\bigl(iA(t_{2n})+\Phi(t_{2n})\bigr)\bigr>.
\]
We are interested in all Wick contractions in the free-field theory, which
represent the contribution of ladder diagrams: in Feynman gauge for the circular loop
we have
\begin{equation}\label{contr}
\langle
\bigl(iA^{a}(t_1)+\Phi^{a}(t_1)\bigr)
\bigl(iA^{b}(t_2)+\Phi^{b}(t_2)\bigr)
\rangle_0
=\frac{g^2\delta^{ab}}{4\pi^2}
\frac{\lvert{\dot x}^{(1)}\rvert\lvert{\dot x}^{(2)}\rvert-{\dot x}^{(1)}
\cdot {\dot x}^{(2)}}{( x^{(1)}-x^{(2)})^2}
=\frac{g^2\delta^{ab}}{8\pi^2}.
\end{equation}
Since each propagator is effectively constant, we can easily perform the sum, and account for the factors of $N$,
by doing the calculation in the zero-dimensional field theory, namely the matrix model (\ref{matrix}).

The above picture has passed many tests along the years but only recently an argument valid
at any order in perturbation theory (and also at non perturbative level) has been proposed
\cite{Pestun:2007rz}: in particular the new construction is gauge-independent and directly
produces the result (\ref{matrix}) for the Wilson loop without referring to Feynman diagrams.
The idea consists in formulating ${\cal N}=4$ SYM on $S^4$: the relevant supersymmetries are generated by conformal
Killing spinors that reduces, in the decompactification limit, to the superconformal spinors (\ref{conformal spinor}).
The partition function of the theory can be computed exactly by deforming the action with a $Q$-exact term and
applying a localization procedure: the path-integral collapses on the zero-modes (on $S^4$
) of some scalar fields of the theory, becoming in this way a simple gaussian matrix model. Quite remarkably gauge fields
play no role in the functional integration, the theory localizing into the (trivial) set of gauge flat-connection
on $S^4$. The circular Wilson loop can be obtained exactly as a $Q$-invariant observable in this construction, thanks
to its BPS property, and therefore localizes into the constant modes as well. Instanton contributions are also argued \cite{Pestun:2007rz}
to decouple completely from the matrix computation and therefore (\ref{matrix}) seems to be exact also at
non perturbative level.

These results can be seen as an explanation of the role of the SUSY invariance in the
cancelations appearing at perturbative level, underlying the birth of the magic matrix integrals: it would
be nice of course to discover more general situations in which exact computations can be performed in this way.
Quite happily a new class of BPS Wilson loops, generalizing the circular loop, has been introduced in \cite{Drukker:2007qr}:
a simple way to understand their construction is to observe that it is possible to pack {\em three} of the six real
scalars into a self-dual tensor
\begin{equation}
\Phi_{\mu\nu}=\sigma^i_{\mu\nu}M^i{}_I\Phi^I\,,
\label{phi-twist}
\end{equation}
and to involve the modified connection
\begin{equation}
A_\mu\to A_\mu+i\Phi_{\mu\nu}x^\nu\,
\end{equation}
in the Wilson loop. The crucial elements in this construction are the tensors
$\sigma^i_{\mu\nu}$: they can be defined by the decomposition of the
Lorentz generators in the anti-chiral spinor representation
($\gamma_{\mu\nu}$) into Pauli matrices $\tau_i$
\begin{equation}
\frac{1}{2}(1-\gamma^5)\gamma_{\mu\nu}
=i\sigma^i_{\mu\nu}\tau_i\,,
\label{gamma-sigma}
\end{equation}
where the projector on the anti-chiral representation is included
($\gamma^5=-\gamma^1\gamma^2\gamma^3\gamma^4$).
The matrix $M^i{}_I$ appearing in (\ref{phi-twist}) is $3\times6$
dimensional and is norm preserving, {\it i.e.}
$M M^{\top}$ is the $3\times 3$ unit matrix (an explicit choice of $M$ is
$M^1{}_1=M^2{}_2=M^3{}_3=1$ and all other entries zero).

These $\sigma$'s are basically the same as 't Hooft's $\eta$
symbols used in writing down instanton solutions, a fact not
surprising because the gauge field is self-dual there. Another, more geometric,
realization of them is in terms of the invariant one-forms on $S^3$
\begin{equation}
\begin{aligned}
\sigma_1^{R,L}& = 2 \left[\pm(x^2 dx^3-x^3 dx^2) +
(x^4 dx^1-x^1 dx^4) \right] \\
\sigma_2^{R,L} &= 2 \left[\pm(x^3 dx^1-x^1 dx^3) +
(x^4 dx^2-x^2 dx^4) \right] \\
\sigma_3^{R,L} &= 2 \left[\pm(x^1 dx^2-x^2 dx^1) +
(x^4 dx^3-x^3 dx^4) \right],
\label{one-forms}
\end{aligned}
\end{equation}
where $\sigma_i^R$ are the right (or left-invariant) one-forms and
$\sigma_i^L$ are the left (or right-invariant) one-forms: explicitly
\begin{equation}
\sigma_i^R=2\sigma^i_{\mu\nu}x^\mu dx^\nu\,.
\label{one-forms-decompose}
\end{equation}

The BPS Wilson loops can then be written in terms of the
modified connection $A_\mu+i\Phi_{\mu\nu}x^\nu$ as
\begin{equation}
W=\frac{1}{N}\,\Tr\,{\cal P}\exp \oint dx^\mu
\left( i A_\mu -\sigma^i_{\mu\nu}x^\nu M^i{}_I\Phi^I \right).
\label{susy-loop}
\end{equation}

We remark that this construction needs the introduction of a length-scale,
as seen by the fact that the tensor (\ref{phi-twist})
has mass dimension one instead of two. The whole procedure
should be consistent therefore when we fix the scale of the Wilson loop.
Actually the operator (\ref{susy-loop}) is supersymmetric only
restricting the loop to be on a three dimensional sphere. This
sphere can be taken embedded in $\mathbb{R}^4$, or coincide with fixed-time slice of
$S^3\times\mathbb{R}$. The authors of \cite{Drukker:2007qr}
have shown that requiring that the supersymmetry variation of these loops vanishes
for arbitrary curves on $S^3$ leads to the two equations
\begin{equation}
\label{susy-fieldtheory}
\begin{aligned}
\gamma_{\mu\nu}\epsilon_1
+i\sigma^i_{\mu\nu}\rho^i\gamma^5\epsilon_0&=0\,,\\
\gamma_{\mu\nu}\epsilon_0
+i\sigma^i_{\mu\nu}\rho^i\gamma^5\epsilon_1&=0\,,
\end{aligned}
\end{equation}
that can be solved consistently: they concluded that for a generic curve on
$S^3$ the Wilson loop preserves $1/16$ of the original supersymmetries.
For special curves, when there are extra relations between the coordinates
and their derivatives, there will be more solutions and the Wilson loops will preserve more
supersymmetry.  A particular interesting case is when the loop lies on a $S^2$: it is possible to
show that these Wilson loops are generically 1/8 BPS and the first perturbative contribution
can be explicitly evaluated \cite{Drukker:2007qr} as
\begin{equation}
<W_{S^2}>=1+g^2N\,\frac{A_1 A_2}{2 A^2}+O(g^4)\,,
\label{g2-result}
\end{equation}
where ${\cal A}$ is the area of the sphere and $A_{1,2}$ are the areas determined by the loop.
This result deserves some comment: first of all we notice that there is no dependence from
the radius of $S^2$, the scale length decoupling consistently with conformal invariance. More intriguingly
we see that the overall result does not depend on the particular shape of the loop, but just on the area
of the two sectors $A_1,A_2$: it suggests a sort of invariance under area preserving transformations.
This fact and the appearance of the peculiar combination $\displaystyle{\frac{A_1 A_2}{ A^2}}$
resemble a similar result for pure Yang-Mills theory on the two-dimensional sphere \cite{Bassetto:1998sr}.
In that case the theory is completely solvable \cite{Migdal:1975zg} and the exact expression for the ordinary Wilson loop is
available \cite{Kazakov:1980zi,Rusakov:1990rs}: restricting the full answer to zero-instanton sector, following the
expansion of \cite{Witten:1992xu}, one obtains
\begin{equation}
<W_0>
=\frac{1}{N}L_{N-1}^1\left(g_{2d}^2\,\frac{A_1 A_2}{A}\right)
\exp\left[-\frac{g_{2d}^2}{2}\,\frac{A_1 A_2}{A}\right]\,,
\label{2d-result}
\end{equation}
where  $L_{N-1}^1(x)$ is a Laguerre polynomial. In the decompactification limit this expression exactly coincides
with the perturbative calculation of \cite{Staudacher:1997kn}, performed by using the light-cone gauge and the Wu-Mandelstam-Liebbrandt prescription
\cite{Wu:1977hi,Mandelstam:1982cb,Leibbrandt:1983pj}, showing that truly non perturbative contributions are not captured in this formulation of the theory. Let us notice that this result is equal to the expectation value of the circular Wilson loop in the gaussian Hermitian matrix model
(\ref{matrix}), after a rescaling of the coupling constant. After identifying the two-dimensional coupling constant $g^2_{2d}$ with
the four-dimensional one through $g^2_{2d}=-g^2/ A$, we see that the first order expansion of (\ref{2d-result}) coincides with
(\ref{g2-result}): the authors of \cite{Drukker:2007qr} have therefore conjectured that the 1/8 BPS Wilson loops constructed on $S^2$
can be computed exactly, leading to the two-dimensional result and claimed more generally the equivalence with the computation of Wilson loop on YM$_2$ on the sphere.

We will try in the following sections to substantiate this conjecture computing higher-order corrections to their results.

\section{Two-loop expansion for supersymmetric loops on $S^2$}
In this section we discuss the expansion at the first two perturbative orders of supersymmetric Wilson loops lying on $S^2$. In order to
perform a quantum analysis in Feynman gauge, we will need to adopt a regularization procedure, since, as we will see, divergent
diagrams could appear in intermediate steps of the computations. We choose the familiar dimensional reduction, consisting in considering
${\cal N}=4$ SYM in $2\omega$ dimensions as a dimensional reduction of ${\cal N}=1$ SYM in ten dimensions. The final results will turn out
nevertheless finite, even in presence of cusps on the contour: we will demonstrate explicitly this property up two loops. In so doing we will derive a compact expression for the $g^4$ contribution of interacting diagrams, that will allow us to perform plain numerical computations in section 4.

The first ingredient in our computations is the effective propagator appearing in the perturbative expansion
of the Wilson loop $i\,$.$\,e$ the analogous of (\ref{contr}), taking into account the explicit couplings of vectors and scalars to
the countour. We shall adopt the short notation $x^\mu_i\equiv x^\mu(t_i)$ and consequently $\dot{x}^\mu_i\equiv \dot{x}^\mu(t_i)$ and
we easily derive
\begin{align}
\Delta^{ab}(t_1,t_2)&=\langle(iA^a_\mu(x_1)\dot x^\mu_1 + |\dot x_1|\Theta^I(t_1)\Phi^a_I(x_1))
(iA^b_\nu(x_2)\dot x^\nu_2 + |\dot x_2|\Theta^J(t_2)\Phi^b_J(x_2))\rangle\nonumber\\
&=\delta^{ab}\frac{\Gamma(\omega-1)}{4\pi^\omega}\frac{|\dot x_1||\dot x_2|\Theta(t_1)\cdot\Theta(t_2) -(\dot x^1\cdot \dot x^2)}{((x_1-x_2)^2)^{\omega-1}}.
\end{align}
The Wilson loops \eqref{susy-loop} are obtained by choosing  $|\dot x_i |\Theta^I(t_i)=
-\sigma^s_{\mu\nu}\dot{x}^\mu_i x^\nu_i M^s{}_I$, where $\sigma^i_{\mu\nu}$ are the 't Hooft
symbol and $M^s{}_I$ is a rectangular matrix, satisfying $M^r{}_I M^s{}_I=\delta^{rs}$ and we have chosen an $S^3$ with unit radius. For
this choice, we find
\be
\begin{split}
|\dot x_1||\dot x_2|
\Theta(t_1)\cdot\Theta(t_2)=&M^r{}_I M^s{}_I
(\sigma^s_{\mu\nu}\dot{x}^\mu_1 x^\nu_1)
(\sigma^r_{\rho\sigma}\dot{x}^\rho_2 x^\sigma_2)=(\sigma^r_{\mu\nu}\dot{x}^\mu_1 x^\nu_1)
(\sigma^r_{\rho\sigma}\dot{x}^\rho_2 x^\sigma_2)\\
=&(\dot{x}_1\cdot \dot{x}_2)(x_1\cdot x_2)-(x_1\cdot\dot{x}_2)(x_2\cdot\dot{x}_1)+\epsilon_{\mu\nu\alpha\beta}
 \dot{x}^\mu_1{x}^\nu_1\dot{x}^\alpha_2{x}^\beta_2,
\end{split}
\ee
where we have exploited the following relation which holds for the 't Hooft symbols
\be
\sigma^i_{\mu\nu}\sigma^i_{\alpha\beta}=\delta_{\mu\alpha}\delta_{\nu\beta}-\delta_{\mu\beta}\delta_{\nu\alpha}+\epsilon_{\mu\nu\alpha\beta}.
\ee
This, in turn, implies that our effective propagator has the simple form
\be
\label{prop1}
\Delta^{ab}(t_1,t_2)\!=\!
\delta^{ab}\frac{\Gamma(\omega-1)}{4\pi^\omega}\frac{
(\dot{x}_1\!\cdot\! \dot{x}_2)[(x_1\!\cdot\! x_2)\!-\!1]\!-\!(x_1 \!\cdot\!\dot{x}_2)(x_2 \!\cdot\!\dot{x}_1)\!+\!\epsilon_{\mu\nu\alpha\beta}
 \dot{x}^\mu_1{x}^\nu_1\dot{x}^\alpha_2{x}^\beta_2 }{((x_1-x_2)^2)^{\omega-1}}.
\ee
Since we shall be  mainly interested in  contours lying  on a unit sphere $S^2$ inside the original $S^3$
the last term in \eqref{prop1} identically vanishes: in this case the four vector $x_1,\dot{x}_1,x_2,\dot{x}_2$
cannot  be in fact linearly independent. We are then left with a reduced propagator of the form
\be
\label{prop2}
\Delta^{ab}(t_1,t_2)\!=\!
\delta^{ab}\frac{\Gamma(\omega-1)}{4\pi^\omega}\frac{
(\dot{x}_1\!\cdot\! \dot{x}_2)[(x_1\!\cdot\! x_2)\!-\!1]\!-\!(x_1 \!\cdot\!\dot{x}_2)(x_2 \!\cdot\!\dot{x}_1) }{((x_1-x_2)^2)^{\omega-1}}.
\ee
In order to investigate the singular behavior of the supersymmetric Wilson loop, it is instructive to study the effective propagator when $t_1$ approaches $t_2$, $i$.$e$. when the two points on the contour are about to collide: it is convenient to rearrange $\Delta^{ab}(t_1,t_2)$
as follows
\be
\begin{split}
\label{propagen}
\!\!\!\!
\Delta^{ab}(t_1,t_2)=&
\delta^{ab}\frac{\Gamma(\omega-1)}{4\pi^\omega}\left[\frac{1}{2}(\dot{x}_1\!\cdot\! \dot{x}_2)((x_1-x_2)^2)^{2-\omega}\!\!\!+\!
\frac{(x_1-x_2)\cdot \dot{x}_2 (x_1-x_2)\cdot\dot{x}_1}{((x_1-x_2)^2)^{\omega-1}}\right].
\end{split}
\ee
Let us consider the case of smooth loops: the above expression \eqref{propagen} is composed by two contributions, that are of
the same order in the coincidence limit. A straightforward Taylor-expansion gives indeed for \eqref{propagen} the following leading behavior
$\Gamma(\omega-1)(|\dot{x}_1|^2)^{3-\omega}(t_1-t_2)^{4-2\omega}$  and it is completely
finite when $1<\omega\le 2$. An analogous result holds for smooth loops with the usual constant coupling, where $\Theta^I$ is a constant
unit vector in $\mathbb{R}^6$, as first shown in \cite{Erickson:2000af}. In this last case divergencies at coinciding points could appear
when considering loops endowed with cusps and are related to the famous "cusp anomaly", as discussed in \cite{Makeenko:2006ds}. We can examine
the non-smooth loop in our case as well: here the situations is more subtle. Let $x_1$  and $x_2$ be the extreme of the propagator approaching the
cusp from the left and the right respectively.  If the cusp is located at $x=x_0$, we can always choose the parametrization of the
contour such that  $x_1=x_0+ t_1 n_1+ O(t_1^2)$ and
$x_2=x_0+ t_2 n_2+ O(t_2^2)$ with $n_1^2=n_2^2=1$ and  $t_1,t_2\ge 0$. Then we find that the leading behavior
of $\Delta^{ab}(t_1,t_2)$ when both $t_1$ and $t_2$ are close to  zero is given just by the second term
\be
\begin{split}
\label{propagen1}
\!\!\!\!
\Delta^{ab}(t_1,t_2)\sim &
\delta^{ab}\frac{\Gamma(\omega-1)}{4\pi^\omega}
\frac{2(1-(n_1\cdot n_2)^2)  t_1 t_2 }{(t_1^2+t_2^2-2 t_1 t_2 (n_1\cdot n_2))^{\omega-1}}.
\end{split}
\ee
A simple power counting argument shows that this object is integrable around $(t_1,t_2)=(0,0)$
for all the values of  $\omega$ less than $3$ and in particular for $\omega=2$. This regular
behavior entails therefore an important difference between the family of loops with the new
scalar couplings and the ones previously considered: no singular contribution is associated here to the presence of the cusp.
In some way the celebrated cusp anomaly appears to be smoothed away at the leading order when considering this
class of supersymmetric Wilson loop. Actually we shall see that this also occurs at two loops and probably it
is true at all orders.

\noindent
We are ready  to list and to analyze the different contributions in the expansion up to the order
$g^4$. We shall perform firstly our analysis for a generic contour and subsequently, in the next
section, we shall consider the specific example of spherical sector whose boundary is two longitudes
of the sphere $S^2$.

At the order $g^2$, since no singular behavior is present at coincident points we can write the relevant integral
directly in four dimensions. Denoting the loop with $C$, we find that
the {\it single-exchange contribution} is given by
\be
\label{single-exch}
W_1(C)
= \frac{g^2 N}{8\pi^2} \oint_C d t_1 dt_2
\frac{
(\dot{x}_1\!\cdot\! \dot{x}_2)[(x_1\!\cdot\! x_2)\!-\!1]\!-\!(x_1 \!\cdot\!\dot{x}_2)(x_2 \!\cdot\!\dot{x}_1) }{(x_1-x_2)^2}\equiv\frac{g^2 N}{8\pi^2}\Sigma_2[C].
\ee

At the order $g^4$, the different contributions are not
separately finite and we have to introduce the regularization procedure.
Firstly,  we shall consider the effect of the one loop correction to the effective propagator
(\ref{prop2}). The relevant diagrams are schematically displayed in fig. \ref{bubble}
and in the following we shall refer to them as  the {\it bubble diagrams}.
\FIGURE[ht]{
\epsfig{file=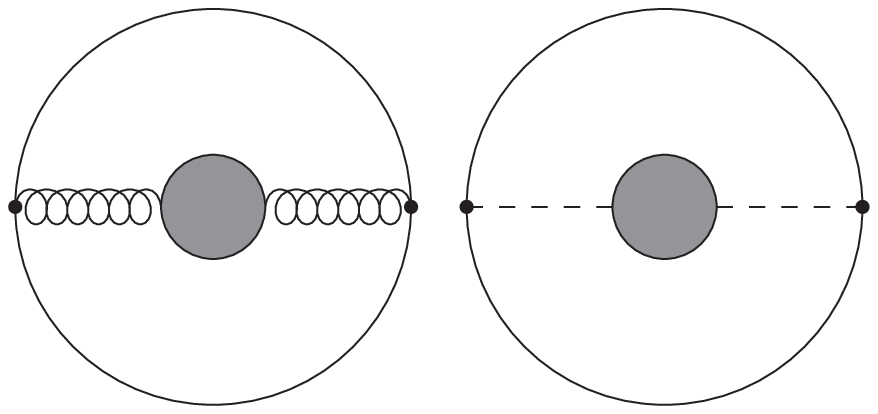,width=4cm }
\caption{\label{bubble} One-loop correction to the gluon and the scalar exchange.}
}
\noindent
The value of the contribution in Feynman gauge can be easily computed with the help of \cite{Erickson:2000af},
where the one-loop correction to the gauge and scalar propagator
has been calculated.  The final result is
\begin{equation}
\begin{split}
&S_2=-g^4 (N^2-1)\frac{\Gamma^2(\omega-1)}
{2^{7}\pi^{2\omega}(2-\omega)(2\omega-3)}\times\\
&\times\oint d\tau_1\, d\tau_2 \frac{(\dot{x}_1\!\cdot\! \dot{x}_2)[(x_1\!\cdot\! x_2)\!-\!1]\!-\!(x_1 \!\cdot\!\dot{x}_2)(x_2 \!\cdot\!\dot{x}_1)
}{\bigl[(x^{(1)}-x^{(2)})^2\bigr]^{2\omega-3} }\equiv\\
&\equiv -g^4 (N^2-1)\frac{\Gamma^2(\omega-1)}
{2^{7}\pi^{2\omega}(2-\omega)(2\omega-3)} \Sigma_\omega[C].
\label{bubbleres}
\end{split}
\end{equation}
Apart from the different power in the denominator, the integrand is identical to \eqref{single-exch}.
The coefficient instead exhibits a pole in $\omega=2$, which keeps track of the divergence in the loop
integration.

\noindent
The next step, at this order, is to investigate the so-called \textit{spider diagrams}, namely the perturbative
contributions coming from the gauge vertex $A^3$ and the scalar-gauge vertex $\phi^2 A$ (see fig. \ref{spider}).  We have to compute
\be
\label{spider1}
S_3=
\frac{g^3}{3 N}\oint d{t}_1 d{t}_2 d{t}_3\eta({t}_1{t}_2{t}_3)
\langle \mathrm{Tr}[\mathcal{A}({t}_1)\mathcal{A}({t}_2)\mathcal{A}({t}_3)]\rangle_0,
\ee
\noindent
where the short notation $\mathcal{A}({t}_\ell)$ stands for the relevant combination
\[
 iA^a_\mu(x_\ell)\dot x^\mu_\ell -\sigma^s_{\mu\nu}\dot{x}^\mu_\ell x^\nu_\ell M^s{}_I \Phi^a_I(x_\ell)
\]
and
\be
\label{spider3}
\eta(t_1,t_2,t_3)=\theta(t_1-t_2)\theta(t_2-t_3)+\mathrm{cyclic\ permutations}.
\ee
After a simple, but tedious computation, in Feynman gauge $S_3$ takes the form
\be
\label{spider4}
\begin{split}
S_3=&\frac{ g^4 (N^2-1)}{4} \oint dt_1 dt_2 dt_3~~\epsilon(t_1,t_2,t_3)\times\\
&\times\left[
(\dot{x}_1\!\cdot\! \dot{x}_3)[(x_1\!\cdot\! x_3)\!-\!1]\!-\!(x_1 \!\cdot\!\dot{x}_3)(x_3 \!\cdot\!\dot{x}_1)\right]\dot{x}^\mu_2\frac{\partial \mathcal{I}_1(x_3-x_1,x_2-x_1)}{\partial x_3^\mu},
\end{split}
\ee
where we have introduced the symbol
\[
\epsilon(t_1,t_2,t_3)=\eta(t_1,t_2,t_3)-\eta(t_2,t_1,t_3),
\]
that is a totally antisymmetric object in the permutations of $(t_1,t_2,t_3)$ and its value is $1$ when $t_1>t_2>t_3$. The
quantity   $\mathcal{I}_1(x,y)$\footnote{$\mathcal{I}_1(x,y)$ is evaluated and its properties  are discussed in details in appendix $B$.}
is defined as the following integral in momentum space
\FIGURE[ht]
{\epsfig{file=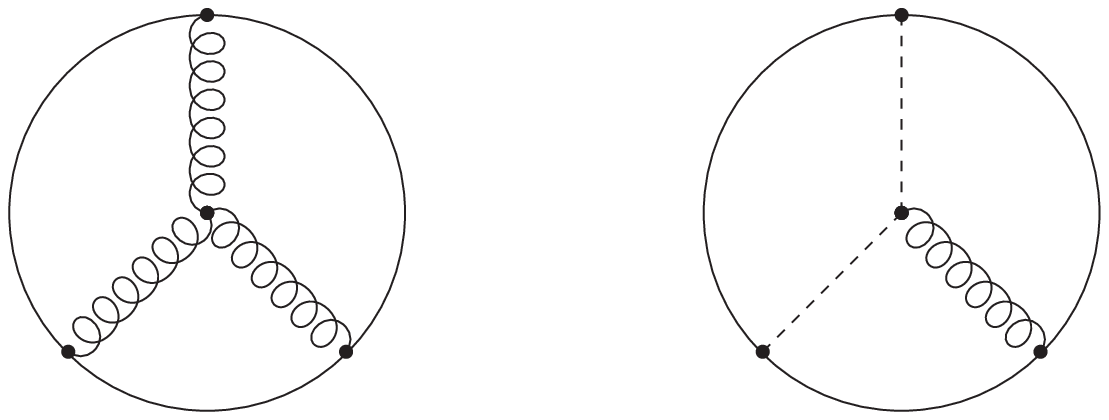,width=7cm}
\caption{ \label{spider} Spider-diagrams: gauge and scalar  contribution}
}\noindent
\be
\label{I1}
\mathcal{I}_1(x,y)\equiv
\int \frac{d^{2\omega} p_1 d^{2\omega} p_2}{(2\pi)^{4\omega}}\frac{e^{i p_1 x+i p_2 y}}{p_1^2
p_2^2 (p_1+p_2)^2}.
\ee
It is quite important, at this point, to understand the potential divergences arising in \eqref{spider4} when $\omega\to 2$: their
appearance originates directly from the integration over the contour. In fact. since the integral \eqref{I1}
is finite and regular for $x,y\ne0$, singularities can only arise in the contour integration when two of the $x_i$ collide.
In that case, a  pole at  $\omega=2$ appears in the expression of
$\mathcal{I}_1(x,y)$:  for $x_1=x_2$ one finds (see app. B)
\be
\mathcal{I}_1(x_3-x_1,0)=\frac{\Gamma^2 (\omega -1)}{(2 \omega-3)(2-\omega)}\frac{1}{64 \pi ^{2 \omega }\left[{(x_1-x_3)^2}\right]^{2\omega-3}}.
\ee
The same behavior occurs when $x_1=x_3$ or $x_2=x_3$ since $\mathcal{I}_1$ is totally symmetric
in the exchange of the $x_i$: we observe three different regions, namely $[(x_1\simeq x_2),\ (x_1\simeq x_3),\ (x_2\simeq x_3)]$,
which are potential sources of divergences. Actually, the situation is better than what one would naively expect:
the true singularity at $\omega\to 2$ appears just in a single region, for the following reasons. One observes that the divergent
behavior at  $x_1=x_3$ becomes integrable  because of the presence of the kinematical pre-factor
$(\dot{x}_1 \cdot  \dot{x}_3)[(x_1 \cdot  x_3) - 1] - (x_1  \cdot \dot{x}_3)(x_3  \cdot \dot{x}_1)$, inherited by the vector/scalar coupling,
which nicely vanishes in this limit. The contribution coming from the region $x_1\simeq x_2$ becomes instead ineffective due to the derivative
with respect to $x_3$, when acting on $\mathcal{I}_1$. The only dangerous singularity appears when $x_3$ approaches $x_2$.

\noindent
A similar pattern for the di\-ver\-gen\-ces was discussed in \cite{Erickson:2000af} for the usual Wilson-Maldacena loop (the loop with constant $\Theta^I$). The authors made the crucial observation that the residual divergence at $x_2\simeq x_3$  is exactly compensated by a contribution coming
from the one-loop correction to the {\it effective} propagator (see fig.~\ref{bubble}, {\it
bubble diagram}). A subtle cancelation among singularities in the contour integration and the loop integration yields
a completely finite result for the Wilson-Maldacena loop at the fourth-order in perturbation theory, in the case of smooth circuits.
This nice conclusion suggests that the divergences appearing in each diagram are indeed gauge artefacts and do not have a physical meaning,
canceling out in the final result. In particular, one could expect that all the diagrams can be made separately finite with a suitable choice of gauge: in appendix $A$, as an example, it is shown that the light-cone gauge does enjoy this property for smooth circuits lying in the plane orthogonal
to light-cone directions.

\noindent
The situation is analogous for the class of supersymmetric Wilson-loop we are considering. Firstly, we shall show that we can explicitly factor
out the divergent part of the {\it spider diagram} and that it has the same form of the {\it bubble} contribution. This can be achieved by rearranging  the original expression \eqref{spider4} for $S_3$  with the help of this trivial identity:
\be
\label{spidertotder}
\begin{split}
0=\frac{ g^4 (N^2-1)}{4} \oint dt_1 dt_2 dt_3~~&\frac{d}{d t_2}\Bigl[\epsilon(t_1,t_2,t_3)\left(
(\dot{x}_1\!\cdot\! \dot{x}_3)((x_1\cdot x_3)-1)-\right.\\
&-\left.\!(x_1 \!\cdot\!\dot{x}_3)(x_3 \!\cdot\!\dot{x}_1)\right) \mathcal{I}_2(x_3-x_2,x_1-x_2)\Bigr].
\end{split}
\ee
The definition and the properties of  the function $\mathcal{I}_2$ are listed in appendix B. With this addition,
the expression can be rearranged by decomposing $S_3$ as the sum of two different contributions, $S_3={\cal A}+{\cal B}$, as follows
\begin{equation*}
\left.\begin{aligned}
\!\!S_3=\!\frac{g^4 (N^2-1)}{4}\!\! \oint& dt_1 dt_2 dt_3\epsilon(t_1,t_2,t_3)\!\left[
(\dot{x}_1\!\cdot\! \dot{x}_3)[(x_1\!\cdot\! x_3)\!-\!1]\!-\!(x_1 \!\cdot\!\dot{x}_3)(x_3 \!\cdot\!\dot{x}_1)\right]\!\times\!\\
&\times\dot{x}^\mu_2\left[\frac{\partial \mathcal{I}_1(x_3-x_1,x_2-x_1)}{\partial x_3^\mu}-\frac{\partial \mathcal{I}_2(x_3-x_1,x_2-x_1)}{\partial x_2^\mu}\right]
\end{aligned}\right\}\!\mathcal{(A)}-
\end{equation*}
\be
\label{spider5}
\left.
\begin{aligned}
-\frac{ g^4 (N^2-1)}{2} \oint &dt_1 dt_3\left(
(\dot{x}_1\!\cdot\! \dot{x}_3)((x_1\cdot x_3)-1)-(x_1 \!\cdot\!\dot{x}_3)(x_3 \!\cdot\!\dot{x}_1)\right)\times\\
&\times \bigl[\mathcal{I}_2(x_3-x_1,x_3-x_1)-\mathcal{I}_2(x_3-x_1,0)\bigr],
\end{aligned}\right\}\mathcal{(B)}
\ee
where we have used
\be
\frac{d}{d t_2} \epsilon(t_1,t_2,t_3)=2( \delta(t_2-t_3)- \delta(t_1-t_2)).
\ee
We start by focussing our attention on ${\cal B}$: it has exactly the same structure of the result $S_2$,
produced by the bubble diagrams, as can be easily inferred looking at the kinematical prefactor. We are led to collect
all these contributions and sum them together. Exploiting the explicit behavior of $I_2(x,y)$ for
$y=x$ and $y=0$, as given in appendix B, we can write the sum of all \textit{bubble-like} contributions
$\mathcal{B}_{\mathrm{tot}}$ as
\be
\label{bubbletot}
\begin{split}
\mathcal{B}_{\mathrm{tot}}=&S_2+\mathrm{{\cal B}}=\frac{g^4 \left(N^2-1\right)}{128 \pi ^{2 \omega-1 }\sin\pi\omega}
\left(\frac{\Gamma(\omega-2)}{\Gamma(3-\omega)}-2\Gamma (2 \omega -4)\right)\Sigma_\omega[C]=\\
=&-
\frac{g^4 \left(N^2-1\right)}{384 \pi ^2}\Sigma_2[C]+O\left((\omega -2)\right).
\end{split}
\ee
In other words, when we sum the term ${\cal B}$ present in \eqref{spider5}
to the original one-loop correction coming from the \textit{bubble diagrams}, we
obtain a completely finite result $\mathcal{B}_{\mathrm{tot}}$, where the pole in
$\omega=2$ has disappeared. Since the contour integration is also finite in this
limit, we can consistently pose $\omega=2$ in \eqref{bubbletot}.

\noindent
The finiteness of \eqref{bubbletot} clearly hints that also the combination
${\cal A}$ appearing in \eqref{spider5} is free of divergencies, as $\omega$ approaches two: this is indeed the case.
In appendix B we show that contribution ${\cal A}$ can be rewritten
as follows, once the derivatives have been explicitly taken
\be
\label{spidersub}
\begin{split}
\mathcal{A}=-&\frac{g^4(N^2-1)\Gamma(2\omega-2)}{128\pi^{{2\omega}}(\omega-1) }
\oint dt_1 dt_2 dt_3\epsilon(t_1,t_2,t_3)\!\left[
(\dot{x}_3\!\cdot\! \dot{x}_1)[(x_1\!\cdot\! x_3)\!-\!1]\!-\right.\\
&-\left.\!(x_1 \!\cdot\!\dot{x}_3)(x_3 \!\cdot\!\dot{x}_1)\right]
\!\!\int_0^1\!\!\!
 d\alpha
{}
\frac{{[\alpha(1-\alpha)]^{\omega-1}\dot{x}_2\cdot(x_1-x_3)~~{_2F_1}(1,2\omega-2;\omega;\xi )}}{(\alpha(x_3-x_2)^2+(1-\alpha) (x_2-x_1)^2)^{2\omega-2}}.
\end{split}
\ee
Here we have denoted with $\xi$ the following combination of the original coordinates
\[
\frac{(x_3-x_1-\alpha(x_2-x_1))^2}{\alpha(x_3-x_2)^2+(1-\alpha) (x_2-x_1)^2},
\]
which appears in the argument of the hypergeometric function ${_2F_1}(1,2\omega-2;\omega;\xi )$.
The integral \eqref{spidersub} is nicely convergent in the limit $\omega\to 2$, in fact by setting
$\omega=2$ we find
\be
\label{spidersub2}
\begin{split}
\!\!\!\!\!\!
\mathcal{A}=&\frac{g^4(N^2-1)}{128\pi^{4} }
\oint dt_1 dt_2 dt_3\epsilon(t_1,t_2,t_3)\!\frac{
(\dot{x}_1\!\cdot\! \dot{x}_3)[(x_1\!\cdot\! x_3)\!-\!1]\!-\!(x_1 \!\cdot\!\dot{x}_3)(x_3 \!\cdot\!\dot{x}_1)}{(x_3-x_1)^2}
\times\\
&\times{\dot{x}_2\cdot(x_3-x_1)}{}
\!\!\int_0^1\!\!\!
 d\alpha
{}
\frac{1}{\alpha(x_3-x_2)^2+(1-\alpha) (x_2-x_1)^2}=\\
=&-\frac{g^4(N^2-1)}{128\pi^{4} }
\oint dt_1 dt_2 dt_3\epsilon(t_1,t_2,t_3)\!\frac{
(\dot{x}_1\!\cdot\! \dot{x}_3)[(x_1\!\cdot\! x_3)\!-\!1]\!-\!(x_1 \!\cdot\!\dot{x}_3)(x_3 \!\cdot\!\dot{x}_1)}{(x_3-x_1)^2}
\times\\
&\ \ \ \ \ \ \ \  \ \ \ \ \ \ \ \ \times{}{}
{}
\frac{\dot{x}_2\cdot(x_3-x_1)}{(x_3-x_2)^2- (x_2-x_1)^2}\log\left(\frac{(x_2-x_1)^2}{(x_3-x_2)^2}\right)
.
\end{split}
\ee
We remark that the original power-like singularity for $x_2\to x_3$ has disappeared and it has been replaced by a milder logarithmic one,
which is integrable both for smooth and cusped loops.

\noindent
We can actually go further and extract from (\ref{spidersub2}) another bubble-like contribution that cancels completely ${\cal B}_{tot}$!
With the help of  the following identity
\be
\begin{split}
&\frac{(x_3-x_1)\cdot\dot{x}_2}{(x_3-x_2)^2-(x_1-x_2)^2}\log\left(\frac{(x_2-x_1)^2}{(x_3-x_2)^2}\right)=
\frac{1}{2}\frac{d}{d t_2} \left[\mathrm{Li}_2\left(1-\frac{(x_2-x_1)^2}{(x_3-x_2)^2}\right)+\right.\\
&\left.+
\frac{1}{2} \left(\log\left[\frac{(x_3-x_2)^2}{(x_3-x_1)^2}\right]\right)^2\right]+
\frac{(x_3-x_2)\cdot\dot{x}_2}{(x_3-x_2)^2}\log\left(\frac{(x_2-x a_1)^2}{(x_3-x_1)^2}\right),
\end{split}
\ee
we can integrate by part \eqref{spidersub2}. We arrive to the following expression
\be
\label{spidersub1}
\begin{split}
\!\!\!\!\!\!
{\mathcal{A}}=&\frac{g^4(N^2-1)}{384\pi^{2} }\,\,\,\Sigma_2[C]
+\\
&+\frac{g^4(N^2-1)}{128\pi^{4} }
\oint dt_1 dt_2 dt_3\epsilon(t_1,t_2,t_3)\!\frac{
(\dot{x}_1\!\cdot\! \dot{x}_3)[(x_1\!\cdot\! x_3)\!-\!1]\!-\!(x_1 \!\cdot\!\dot{x}_3)(x_3 \!\cdot\!\dot{x}_1)}{(x_3-x_1)^2}
\times\\
&\ \ \ \ \ \ \ \  \ \ \ \ \ \ \ \ \times{}{}
{}
\frac{(x_3-x_2)\cdot\dot{x}_2}{(x_3-x_2)^2}\log\left(\frac{(x_2-x_1)^2}{(x_3-x_1)^2}\right).
\end{split}
\ee
We see that the first term exactly cancels $\mathcal{B}_{tot}$ and the only surviving contribution from the {spider and the bubble}
diagrams can be written as a relatively simple convergent integral
\be
\label{spiderfin}
\begin{split}
\!\!\!\!\!\!
\mathcal{I}_{\textrm{tot}}=&\frac{g^4(N^2-1)}{128\pi^{4} }
\oint dt_1 dt_2 dt_3\epsilon(t_1,t_2,t_3)\!\frac{
(\dot{x}_1\!\cdot\! \dot{x}_3)[(x_1\!\cdot\! x_3)\!-\!1]\!-\!(x_1 \!\cdot\!\dot{x}_3)(x_3 \!\cdot\!\dot{x}_1)}{(x_3-x_1)^2}
\times\\
&\ \ \ \ \ \ \ \  \ \ \ \ \ \ \ \ \times{}{}
{}
\frac{(x_3-x_2)\cdot\dot{x}_2}{(x_3-x_2)^2}\log\left(\frac{(x_2-x_1)^2}{(x_3-x_1)^2}\right).
\end{split}
\ee
It is remarkable that this expression holds for any kind of loop on $S^2$, both cusped and smooth, and being free of divergencies
is amenable, if necessary, to a plain numerical evaluation, once the contour is specified. An analogous representation for planar contours with a constant $\Theta^I$ coupling is given in Appendix A: in the circular case $I_{tot}$ is easily seen to vanish by simple symmetry arguments,
recovering without tears the result of \cite{Erickson:2000af}.

\noindent
This is not of course the end of story: we have still to consider the \textit{double-exchange} diagrams to the perturbative expansion of the Wilson loop, namely we have to analyze the contribution
\be
\frac{g^4}{N}\oint_C d t_1 d t_2  d t_3  d t_4\theta(t_1-t_2)\theta(t_2-t_3)\theta(t_3-t_4)
\langle \mathrm{Tr}[\mathcal{A}(t_1) \mathcal{A}(t_2) \mathcal{A}(t_3) \mathcal{A}(t_4)]\rangle_0.
\ee
Recalling that the effective propagator has the color structure $\Delta^{ab}(t_1,t_2)=\delta^{ab}\Delta(t_1,t_2)$, the relevant Green function can
be written as
\be
\begin{split}
&\langle \mathrm{Tr}[\mathcal{A}(t_1) \mathcal{A}(t_2) \mathcal{A}(t_3) \mathcal{A}(t_4)]\rangle_0
=\frac{1}{2}\mathrm{Tr}([T^b, T^a][T^b,T^a])\Delta(t_1,t_3) \Delta(t_2,t_4)+\\
&+\mathrm{Tr}(T^aT^a T^b T^b) \left[\Delta(t_1,t_2) \Delta(t_3,t_4)+\Delta(t_1,t_3) \Delta(t_2,t_4)+\Delta(t_1,t_4) \Delta(t_2,t_3)\right].
\end{split}
\ee
The term multiplying $\mathrm{Tr}(T^aT^a T^b T^b)$ is symmetric in the exchange of all the $t_i$ and therefore is insensitive to the path-ordering. It simply yields $1/2$ the square of
the \textit{single-exchange} contribution
\be
\frac{1}{2}\left(\frac{g^2 N}{8\pi^2} \oint_C d t_1 dt_2
\frac{
(\dot{x}_1\!\cdot\! \dot{x}_2)[(x_1\!\cdot\! x_2)\!-\!1]\!-\!(x_1 \!\cdot\!\dot{x}_2)(x_2 \!\cdot\!\dot{x}_1) }{(x_1-x_2)^2}\right)^2.
\ee
This simple manipulation expresses the trivial exponentiation of the so-called abelian part of the Wilson loop. The remaining contribution,
which is proportional to  $\mathrm{Tr}([T^b, T^a][T^b,T^a])$, is usually called the \textit{maximally non-abelian part} and
it is the new ingredient in the \textit{double-exchange} contribution. We are left to compute the integral
\be
\label{maxnab}
\begin{split}
&-\frac{g^4(N^2-1)\Gamma^2(\omega-1)}{64\pi^{2\omega}}
\oint_C d t_1 d t_2  d t_3  d t_4\theta(t_1-t_2)\theta(t_2-t_3)\theta(t_3-t_4)\times\\
&\times\frac{
(\dot{x}_1\!\cdot\! \dot{x}_3)[(x_1\!\cdot\! x_3)\!-\!1]\!-\!(x_1 \!\cdot\!\dot{x}_3)(x_3 \!\cdot\!\dot{x}_1) }{((x_1-x_3)^2)^{\omega-1}}
\times
\frac{
(\dot{x}_4\!\cdot\! \dot{x}_2)[(x_4\!\cdot\! x_2)\!-\!1]\!-\!(x_4 \!\cdot\!\dot{x}_2)(x_2 \!\cdot\!\dot{x}_4) }{((x_4-x_2)^2)^{\omega-1}}.
\end{split}
\ee
This contribution is of course finite and we can set safely again $\omega=2$.
\noindent

\section{The cusped loop on $S^2$}
In the present section we will provide
a fourth-order evidence that the supersymmetric Wilson loops lying on $S^2$ are actually
equivalent to the usual, non-supersymmetric Wilson loops of Yang-Mills theory on a 2-sphere in the Wu-Mandelstam-Leibbrandt
prescription \cite{Wu:1977hi,Mandelstam:1982cb,Leibbrandt:1983pj}, as conjectured in \cite{Drukker:2007qr} on the basis of
a one-loop calculation. We have not been able to show this equivalence in general: one should compute
(\ref{spiderfin}) and (\ref{maxnab}) for a generic contour on $S^2$ and compare the total result with the expansion of (\ref{2d-result}) at order $g^4$ (the abelian part of the Wilson loop being trivially recovered). This task seems particulary difficult, especially because we do not see any simple way in which (\ref{spiderfin}) and (\ref{maxnab}) could generate something proportional to $(\Sigma_2[C])^2$. In the parent circular case, that corresponds to a contour winding the equator of $S^2$, two obvious simplifications appear: the vanishing of (\ref{spiderfin}) and the constant behavior of the effective propagator, that allows an easy computation of (\ref{maxnab}). For a generic loop on $S^2$  both properties seem to disappear, at least in Feynman gauge, and the matrix model result could be recovered only through a delicate interplay among interacting and double-exchange contributions. We are led therefore to check, as first instance, the conjecture against a particular class of loops, for which the calculation of (\ref{spiderfin}) and (\ref{maxnab}) is relatively easy. We will focus on a particular family of $1/4$ BPS Wilson loops that can be obtained as follows.
Consider a loop made of two arcs of length $\pi$ connected at an arbitrary angle $\delta$, {\it i.e.} two longitudes on the two-sphere: an explicit parametrization is
\be
\label{Contour}
x(t)=\left\{\begin{array}{l}
       \left(- \frac{2 t}{1+t^2},~ 0,\frac{1-t^2}{1+t^2}\right) \ \ \ \ \ \ \ \ \ \ \ \ \ \ \ \ \ \ \ \  \mathrm{for} \ \ \ \  -\infty< t\le 0 \\
       \\
       \left( \frac{2 t}{1+t^2} \cos\delta,~ \frac{2 t}{1+t^2} \sin\delta,~\frac{1-t^2}{1+t^2}\right) \ \ \ \  \mathrm{for} \ \ \ \ \ \ \ \ \   0\le t<\infty
     \end{array}\right.
\ee
This path starts from the south pole of the sphere $(0,0,-1)$  for $(t=-\infty)$. When $t$ increases, we move along a meridian $\phi=0$
up to the north pole  $(0,0,1)$, which is reached for $(t=0)$. From the north pole, we move back to the south pole  along the meridian
$\phi=\delta$ and we again reach the south pole when $t=+\infty$. In other words,
this path is the border of a spherical sector whose angular width is given by $\delta$.  Notice that our parametrization for the contour is nothing
else but its stereographic projection on the plane. After this projection, our contour appears as an infinite angular sector (see fig. \ref{figspicchio1}). We will call our contour the $wedge$.
\FIGURE{\epsfig{file=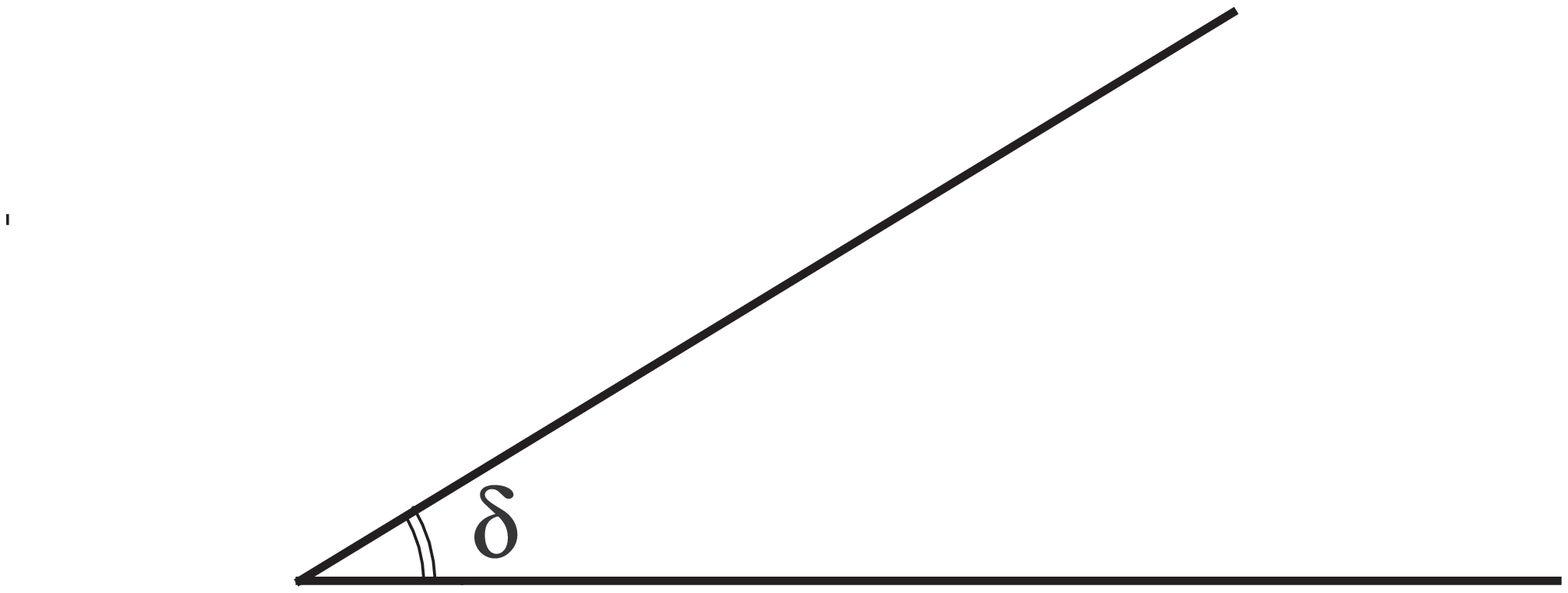,width=9cm}
\caption{ \label{figspicchio1}Stereographic Projection of the \textit{``wedge''}.}}

\noindent
Let us start by discussing, as a warm up, the lowest order contribution. For this kind of loop the \textit{single exchange} splits in  three sub-diagrams:
\FIGURE{
\epsfig{file=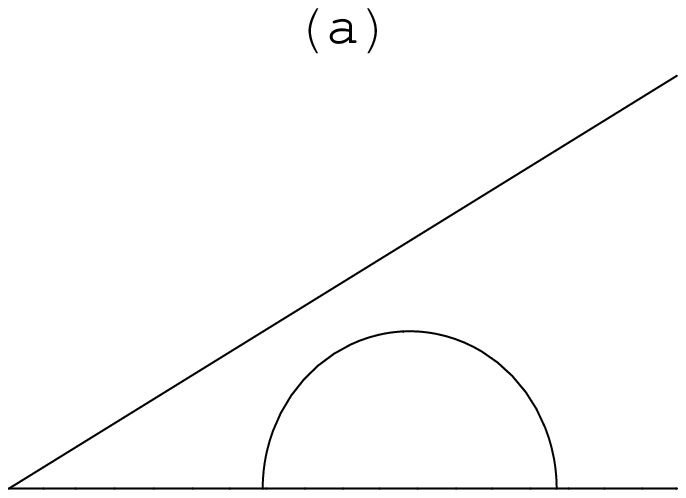,width=4cm}\ \ \ \epsfig{file=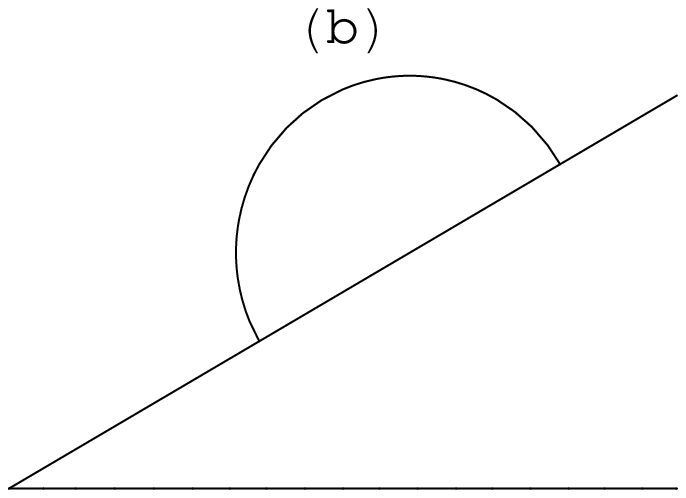,width=4cm}\ \ \ \epsfig{file=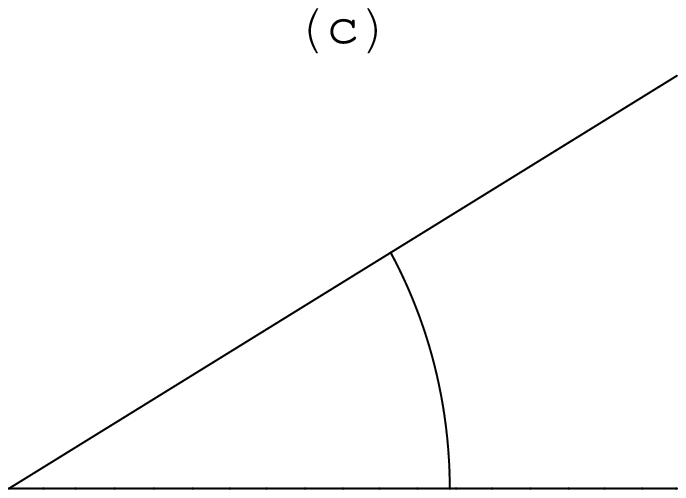,width=4cm}
\caption{\label{SEDs} \textit{Single-exchange} diagrams}
}

\noindent
The diagram $(a)$ and $(b)$ are equal, since we cannot distinguish the two longitudes. We have that
\be
(a)+(b)=2(a)= \frac{g^2 N}{2\pi^2} \int_{-\infty}^0 d t_1\int^{t_1}_{-\infty} dt_2
\frac{1}{\left(t_1^2+1\right) \left(t_2^2+1\right)}
=\frac{g^2 N}{16}.
\ee
The diagram $(c)$ is given by
\be
\begin{split}
(c)
=-\frac{g^2 N}{4\pi^2}&\int^{\infty}_0 dt_1\int^{\infty}_{0} d t_2
\frac{-2 {t_1} {t_2}+\left({t_1}^2+{t_2}^2\right) \cos (\delta )}{ \left({t_1}^2+1\right) \left({t_2}^2+1\right) \left({t_1}^2+{t_2}^2-2
   {t_2}  {t_1} \cos (\delta )\right)},
\end{split}
\ee
where we have performed the change of variable $t_2\mapsto -t_2$. Next we pose  $t_1= t_2 w$ and we integrate over $t_2$.
Then we get\footnote{The integral \eqref{F8} can be computed with the Residue theorem applied to the function
\[
\frac{ (\left(w^2  +1\right) \cos (\delta )-2 w)}{  \left( w^2 +1-2
    w \cos (\delta )\right)}\frac{1}{w^2-1}\left(\frac{i \log ^2(w)}{4 \pi }+\frac{\log (w)}{2}\right)
\]
for a contour that encircles the cut of the logarithm. The cut is taken along the positive real axis.}
\be
\begin{split}
\label{F8}
(c)=-\frac{g^2 N}{4\pi^2}&\int^{\infty}_0 d w \frac{\log(w)}{w^2-1}\frac{ (\left(w^2  +1\right) \cos (\delta )-2 w)}{  \left( w^2 +1-2
    w \cos (\delta )\right)}=-\frac{g^2 N}{4\pi^2}\left(\frac{\pi ^2}{4}-\frac{1}{2} (2 \pi -\delta ) \delta\right).
\end{split}
\ee
Summing the three different contributions, we find the first-order contribution
\be
W_1=(a)+(b)+(c)=\frac{g^2 N}{8\pi^2} (2 \pi -\delta ) \delta,
\ee
consistently with the general result of \cite{Drukker:2007qr} and the related conjecture,once one notes that $A_1A_2/A^2=\delta(2\pi-\delta)/4\pi^2$.

The next step is to tackle the double-exchange diagrams, as first contributions at order $g^4$. Since the \textit{abelian } part of these diagrams is given by $1/2$ the square of the contribution of order $g^2$, as we have seen in the previous section, we shall focus our attention only to the maximally non-abelian part. The relevant diagrams can be separated in three different
families, according to the number of propagators with both ends  on the same edge of the circuit.
\FIGURE{
\epsfig{file=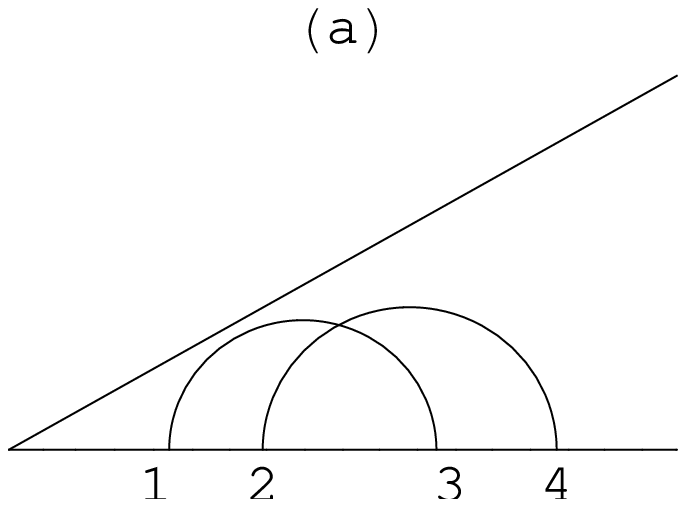,width=4.5cm}\  \ \ \ \epsfig{file=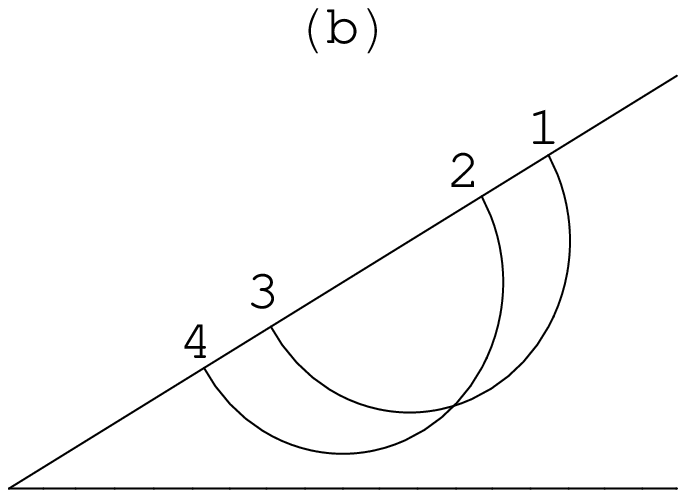,width=4.5cm}
\caption{\label{DESfam(I)} Double exchange diagrams: type (I) }}
To begin with, we have the case of diagrams of fig. \ref{DESfam(I)}. The contributions of diagram (Ia)
and (Ib) are equal. Their value is
\be
\begin{split}
\textrm{(I)}=&\textrm{(Ia)}+\textrm{(Ib)}=2 \textrm{(Ib)}=\\
=&-\frac{g^4(N^2-1)}{8\pi^{4}}\int_0^\infty\!\!\!\! d t_1\! \int_0^{t_1}\!\!\!\! d t_2\!  \int_0^{t_2}
\!\!\!\! d t_3\!  \int_0^{t_3}\!\!\!\! d t_4
\frac{1}{\left({t_1}^2+1\right) \left({t_2}^2+1\right)
   \left({t_3}^2+1\right) \left({t_4}^2+1\right)}=\\
=&-\frac{g^4(N^2-1)}{3072}.
\end{split}
\ee
Consider now the second family of diagrams represented in fig. \ref{DESfam(II)}. Again the two
diagrams are equal and we can write
\be
\begin{split}
\textrm{(II)}=&\textrm{(IIc)}+\textrm{(IId)}=2 \textrm{(IIc)}=\\
=&\frac{g^4(N^2-1)}{8\pi^{4}}\int_0^\infty\!\!\!\! d t_1\! \int_{-\infty}^{0}\!\!\!\! d t_2\!  \int_{-\infty}^{t_2}
\!\!\!\! d t_3\!  \int_{-\infty}^{t_3}\!\!\!\! d t_4
\mbox{\small $\frac{\cos (\delta ) t_1^2+2 t_3 t_1+\cos (\delta ) t_3^2}{ \left(t_1^2+1\right) \left(t_2^2+1\right)
   \left(t_3^2+1\right) \left(t_1^2+2 \cos (\delta ) t_3
   t_1+t_3^2\right) \left(t_4^2+1\right)}$}=\\
=&\frac{g^4(N^2-1)}{8\pi^{4}}\int_0^\infty\!\!\!\! d t_1\! \int_{-\infty}^{0}\!\!\!\! d t_3\!  \int^{0}_{t_3}
\!\!\!\! d t_2\!  \int_{-\infty}^{t_3}\!\!\!\! d t_4
\mbox{\small $\frac{\cos (\delta ) t_1^2+2 t_3 t_1+\cos (\delta ) t_3^2}{ \left(t_1^2+1\right) \left(t_2^2+1\right)
   \left(t_3^2+1\right) \left(t_1^2+2 \cos (\delta ) t_3
   t_1+t_3^2\right) \left(t_4^2+1\right)}$}.
\end{split}
\ee
The integration over $t_2$ and $t_4$ are trivial and can be performed analytically. We have
\be
\begin{split}
\mathrm{(II)}
=&\frac{g^4(N^2-1)}{8\pi^{4}}\int_0^\infty\!\!\!\! d t_1\! \int_{-\infty}^{0}\!\!\!\! d t_3
 \frac{\tan ^{-1}\left(\frac{1}{t_3}\right) \tan ^{-1}\left(t_3\right)(\cos (\delta ) (t_1^2+t_3^2)+2 t_3 t_1)}{ \left(t_1^2+1\right)
   \left(t_3^2+1\right) \left(t_1^2+2 \cos (\delta ) t_3
   t_1+t_3^2\right) }\equiv\\
\equiv& \frac{g^4(N^2-1)}{8\pi^{4}} {\cal D}_1.
\end{split}
\ee
\FIGURE{
\epsfig{file=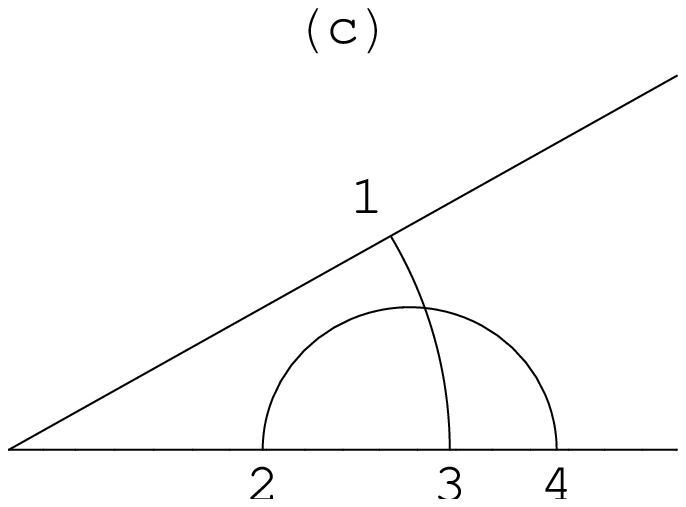,width=4.5cm}\  \ \ \ \epsfig{file=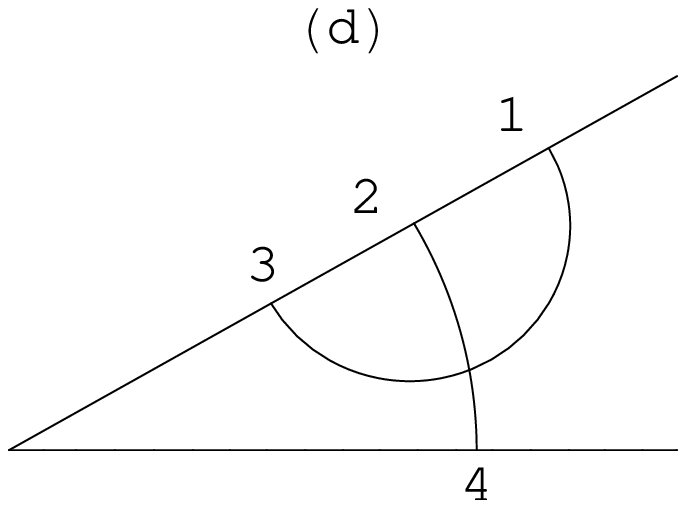,width=4.5cm}
\caption{\label{DESfam(II)} Double exchange diagrams: type (II) }}
We could now perform the integration over $t_1$ since the integrand is a rational function of this variable, but this
is not particularly convenient. We would end up indeed with a function of $t_1$ that we cannot integrate analytically,
but only numerically. For this reason, we start our numerical analysis already at the level of  ${\cal D}_1$.
The result as a function of $\delta$ is given in fig. \ref{DE1}.
\FIGURE{
\epsfig{file=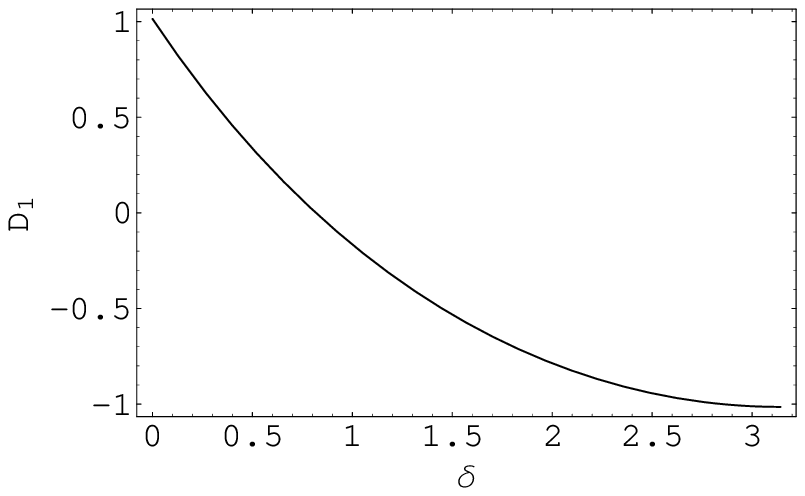,width=9cm}
\caption{\label{DE1} Plot of ${\cal D}_1$ as a function of $\delta$ in the range $[0,\pi]$.
We focus our attention to this interval because all the integrals possess the symmetry $\delta
\mapsto 2\pi -\delta$. }}
\noindent
We consider finally the last diagram contributing to the double-exchange. It is schematically drawn
fig. \ref{DESfam(III)}. The actual integral to evaluate for this diagram is given by
\be
\begin{split}
&\mathrm{(III)}=-\frac{g^4(N^2-1)}{16\pi^{4}}\int_0^{\infty}\!\!\!\! d t_1\!\int^{\infty}_0 \!\!\!\!d t_4\!
\int_0^{t_1}\!\!\!\! d t_2\!
\int^{t_4}_0
\!\!\!\! d t_3  \times\\
&\times\mbox{$\frac{ \left(\left({t_1}^2+{t_3}^2\right) \cos (\delta
   )-2 {t_1} {t_3}\right)
   \left(\left({t_2}^2+{t_4}^2\right) \cos (\delta )-2
   {t_2} {t_4}\right)}{\left({t_1}^2+1\right)
   \left({t_2}^2+1\right) \left({t_3}^2+1\right)
   \left({t_4}^2+1\right) \left({t_1}^2-2 {t_3} \cos
   (\delta ) {t_1}+{t_3}^2\right) \left({t_2}^2-2
   {t_4} \cos (\delta ) {t_2}+{t_4}^2\right)} $}\equiv -\frac{g^4(N^2-1)}{16\pi^{4}} {\cal D}_2,
\end{split}
\ee
\FIGURE{
\epsfig{file=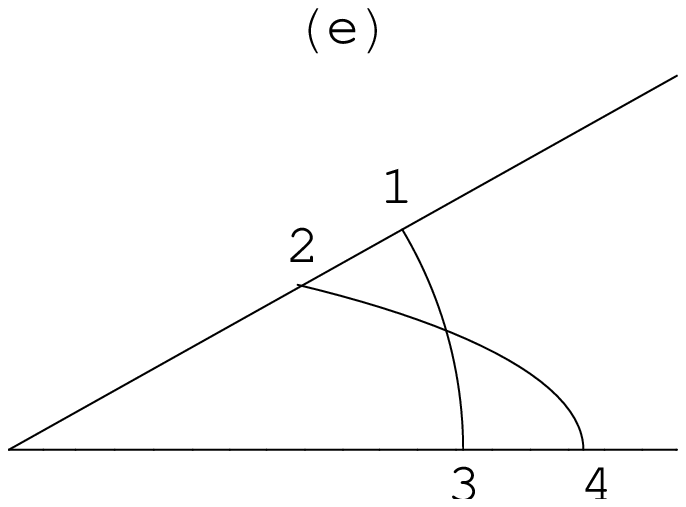,width=5.5cm}
\caption{\label{DESfam(III)}Double-exchange diagrams: type (III) }}
\noindent
where we have performed the following change of  variables $t_3\mapsto -t_3$ and $t_4\mapsto -t_4$
and then we have rearranged the order of the different integrations. In this form, the integration
over $t_2$ and $t_3$ can be performed analytically, while the residual two integrations can
be done numerically. The final result for ${\cal D}_2$ is plotted in fig. \ref{DE2}
\FIGURE[htb]{
\epsfig{file=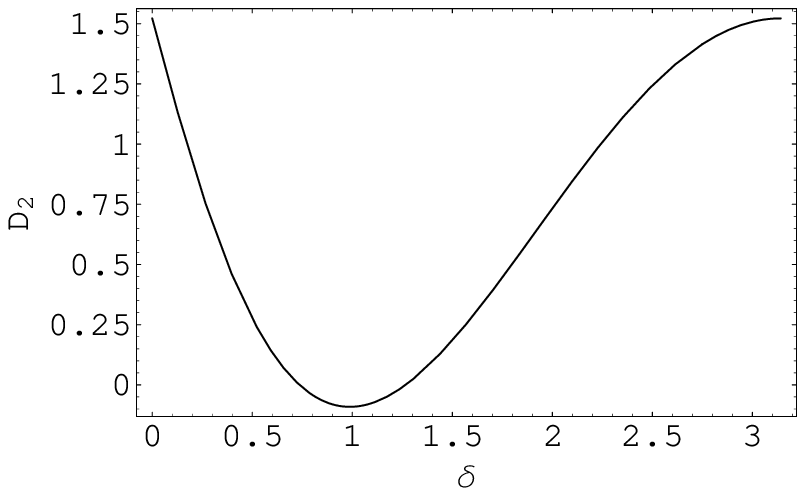,width=9cm}
\caption{\label{DE2} Plot of ${\cal D}_2$ as a function of $\delta$ in the range $[0,\pi]$.
 }}.
\FIGURE{
\epsfig{file=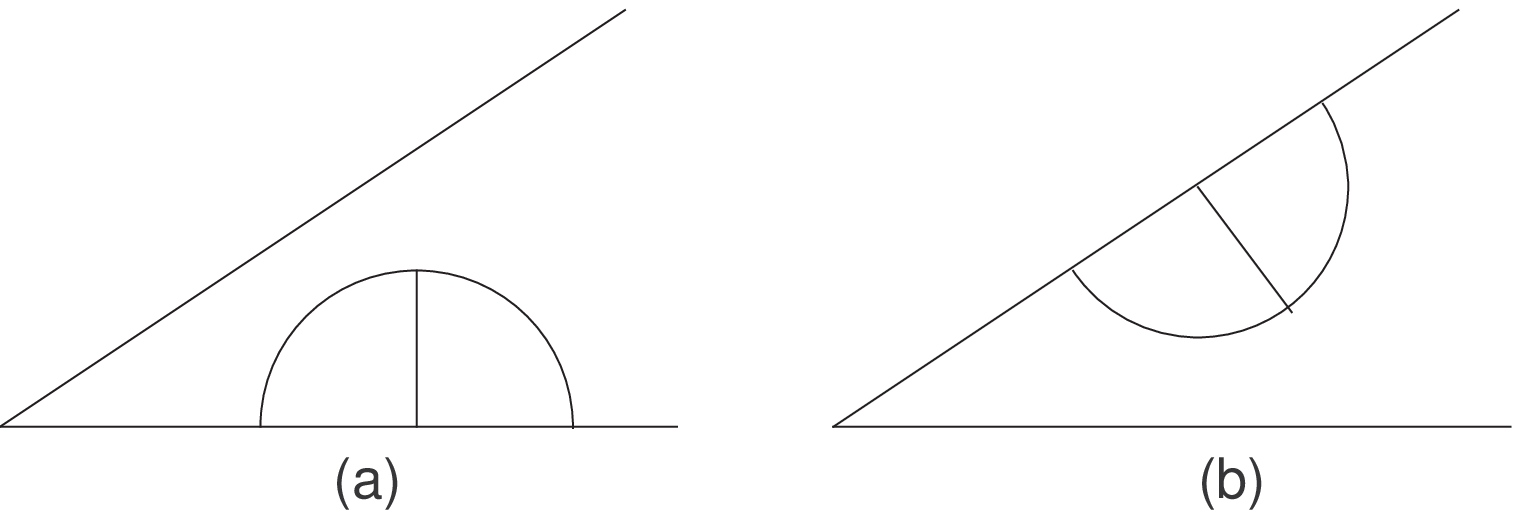,width=9cm}
\caption{\label{samedge} Diagrams of type (A) }}

\noindent
Having evaluated the double-exchange diagrams, we are left to consider the effective contribution due to the interactions and summarized in
the result \eqref{spiderfin}, found in  Section 3. In order to write the actual integrals
we have to compute, we distinguish two cases: (A) when the legs of the vertex are attached
to the same edge of the circuit (see fig. \ref{samedge} ) and (B) when the  legs of the vertex are  not attached
to the same edge of the circuit (see figs. \ref{Intera1}, \ref{Intera2} ). The diagrams belonging to the family (A) vanish. To convince the reader let us consider , for example, the first of the two diagrams that it is given by
\be
\frac{g^4(N^2-1)}{64\pi^{4} }\!\!\!\int_{-\infty}^0\!\!\!\!\!\!\!d t_1\int_{-\infty}^0\!\!\!\!\!\!\!d t_2\int_{-\infty}^0\!\!\!\!\!\! \!d t_3
\epsilon(t_1,t_2,t_3)\frac{ ({t_2} {t_3}+1) \log
   \left(\frac{({t_1}-{t_2})^2
   \left({t_3}^2+1\right)}{\left({t_2}^2+1\right)
   ({t_1}-{t_3})^2}\right)}{\left({t_1}^2+1\right)
   \left({t_2}^2+1\right) ({t_2}-{t_3})
   \left({t_3}^2+1\right)}.
\ee
This integral is equal to zero because the integrand is antisymmetric in the interchange $(t_2,t_3)$, while
the integration region is symmetric. We are finally left with the diagrams belonging to the family (B). We have six diagrams:
(1) three with two legs of the vertex attached to first edge of the spherical wedge and (2) three with two legs
attached to second edge. The contribution of the two classes is equal because our loop is symmetric under reflection
with respect the longitude $\phi=\delta/2$. We shall consider the first class only and we will multiply
the result by two. Then the total contribution of the interaction is given by the following integral
\FIGURE[ht]{
\epsfig{file=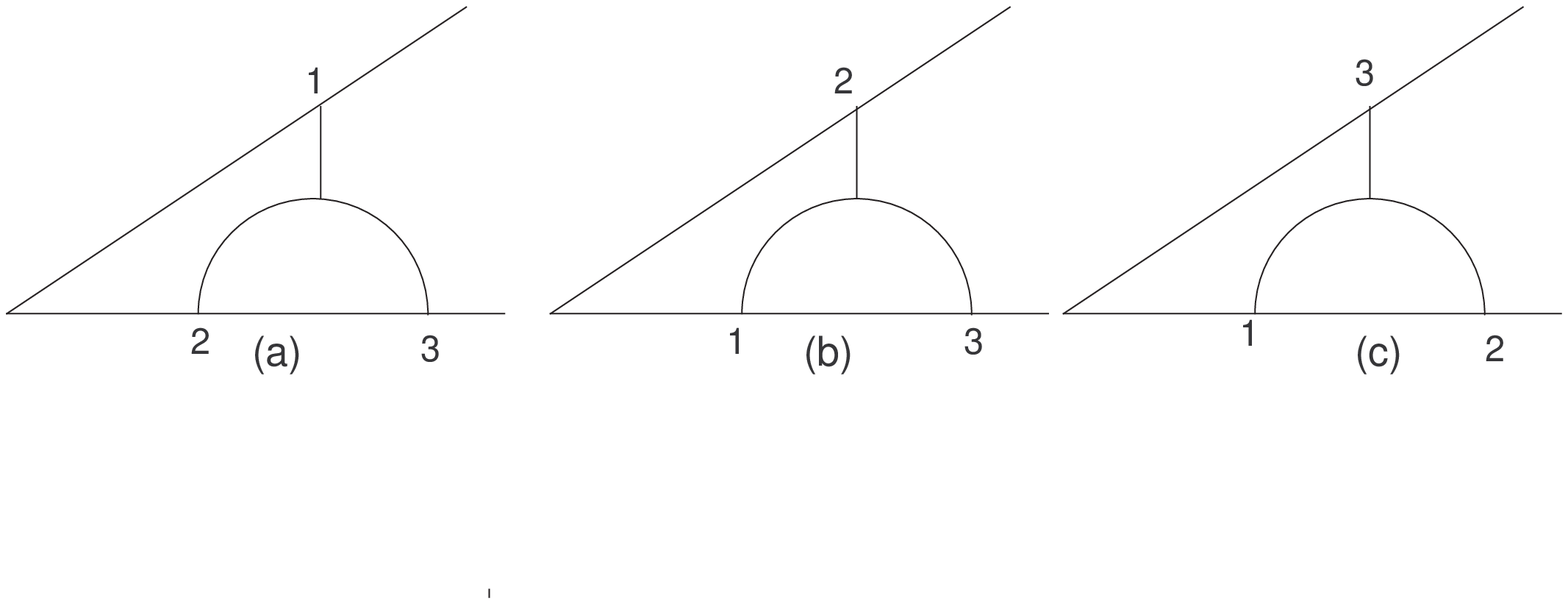,width=9cm}
\caption{\label{Intera1} Diagrams of type (B): set(1)}}
\FIGURE[ht]{
\epsfig{file=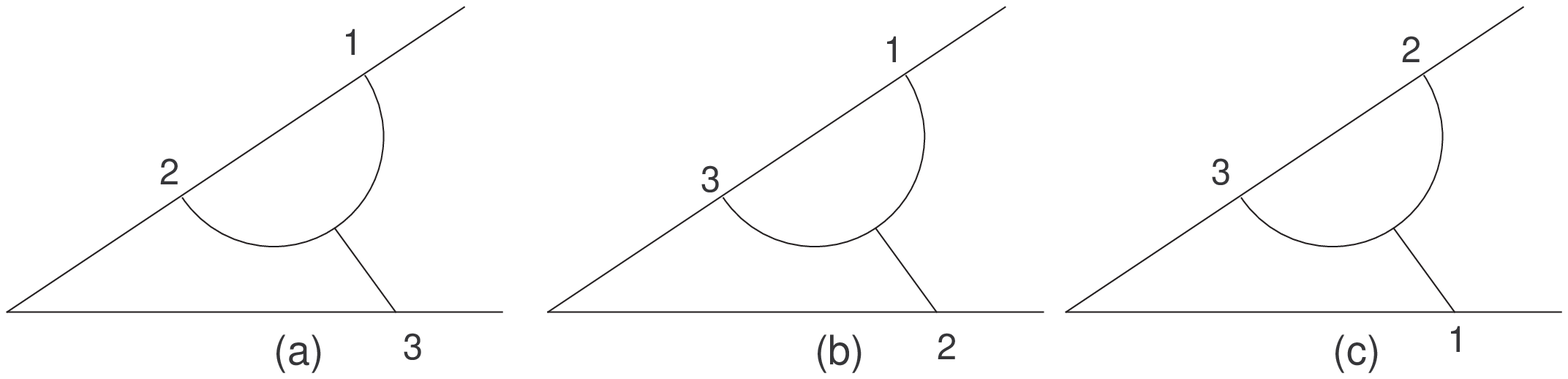,width=9cm}
\caption{\label{Intera2} Diagrams of type (B): set (2)}}
\be
\begin{split}
\mathcal{I}_{tot}=&\frac{g^4(N^2-1)}{64\pi^{4} }
\int_0^\infty \!\!\!\!\!\!dt_1 dt_2 dt_3 \left[
( \text{sgn}({t_3}-{t_2})\mathrm{V}_1+  \text{sgn}({t_1}-{t_3})\mathrm{V}_2 + \text{sgn}({t_2}-{t_1})\mathrm{V}_3\right]\equiv\\
\equiv&\frac{g^4(N^2-1)}{64\pi^{4} } \mathrm{I}_{tot},
\end{split}
\ee
where
\be
\begin{split}
\mathrm{V}_1=&
\mbox{$
\frac{2 ({t_2} {t_3}+1)
   \left(\left({t_1}^2+{t_3}^2\right) \cos (\delta )-2
   {t_1} {t_3}\right) \log
   \left(\frac{\left({t_3}^2+1\right) \left({t_1}^2-2
   {t_2} \cos (\delta )
   {t_1}+{t_2}^2\right)}{\left({t_2}^2+1\right)
   \left({t_1}^2-2 {t_3} \cos (\delta )
   {t_1}+{t_3}^2\right)}\right)
  }{\left({t_1}^2+1\right)
   \left({t_2}^2+1\right) ({t_3}-{t_2})
   \left({t_3}^2+1\right) \left({t_1}^2-2 {t_3} \cos
   (\delta ) {t_1}+{t_3}^2\right)}$}\\
\mathrm{V}_2=&\mbox{$
\frac{2 \left({t_2}
   \left({t_3}^2-1\right)-\left({t_2}^2-1\right) {t_3}
   \cos (\delta )\right) \log
   \left(\frac{\left({t_3}^2+1\right) \left({t_1}^2-2
   {t_2} \cos (\delta )
   {t_1}+{t_2}^2\right)}{\left({t_2}^2+1\right)
   ({t_3}-{t_1})^2}\right)}{\left({t_1}^2+1\right)
   \left({t_2}^2+1\right) \left({t_3}^2+1\right)
   \left({t_2}^2-2 {t_3} \cos (\delta )
   {t_2}+{t_3}^2\right)}$}\\
\mathrm{V}_3=&
\mbox{$
\frac{2 \left(\left({t_2}^2-1\right) {t_3} \cos (\delta
   )-{t_2} \left({t_3}^2-1\right)\right)
   \left(\left({t_1}^2+{t_3}^2\right) \cos (\delta )-2
   {t_1} {t_3}\right) \log
   \left(\frac{({t_2}-{t_1})^2
   \left({t_3}^2+1\right)}{\left({t_2}^2+1\right)
   \left({t_1}^2-2 {t_3} \cos (\delta )
   {t_1}+{t_3}^2\right)}\right)
  }{\left({t_1}^2+1\right)
   \left({t_2}^2+1\right) \left({t_3}^2+1\right)
   \left({t_1}^2-2 {t_3} \cos (\delta )
   {t_1}+{t_3}^2\right) \left({t_2}^2-2 {t_3} \cos
   (\delta ) {t_2}+{t_3}^2\right)}$}.
\end{split}
\ee
If we expand the logarithms in this expression, we can always perform one integration
analytically: the argument of each logarithm always depends only on two of
the three variables. It turns out that one of the three integrations always reduces to find
the primitive of a rational function. As we have done before, the remaining two integrations can
be easily performed numerically and the result is given in fig. \ref{Inte}
\FIGURE{
\epsfig{file=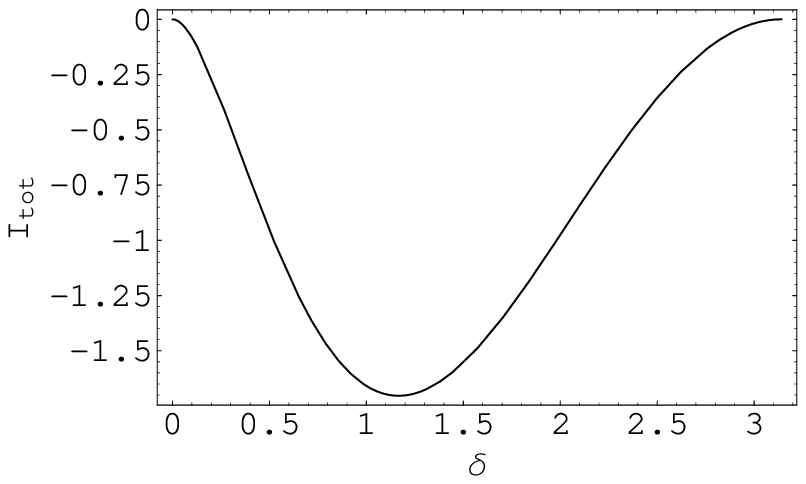,width=9cm}
\caption{\label{Inte} Plot of $\mathrm{I}_{tot}$ as a function of $\delta$ in the range $[0,\pi]$.
 }}.
We can finally collect all the results to obtain the maximal nonabelian contribution at order $g^4$:
\be
\begin{split}
W_2^{mnb}=&-\frac{g^4(N^2-1)}{3072}+\frac{g^4(N^2-1)}{8\pi^{4}} {\cal D}_1-\frac{g^4(N^2-1)}{16\pi^{4}} {\cal D}_2
+\frac{g^4(N^2-1)}{64\pi^{4} } \mathrm{I}_{tot}=\\
=&-\frac{g^4(N^2-1)}{8\pi^4}\left(\frac{\pi^4}{384}-{\cal D}_1+\frac{1}{2}{\cal D}_2-\frac{1}{8}\mathrm{I}_{tot} \right)\equiv
-\frac{g^4(N^2-1)}{8\pi^4} R.
\end{split}
\ee
The plot for $R$ is given in fig. \ref{fin}. We can easily perform a fit of the numerical result $R$
with a polynomial of the following form $P(\delta)= c_0 \delta^2(2 \pi-\delta)^2$. This
particular dependence is necessary in order to be in agreement with the conjectured relation with the zero-instanton sector of
pure Yang-Mills theory on the sphere.
\FIGURE{
\epsfig{file=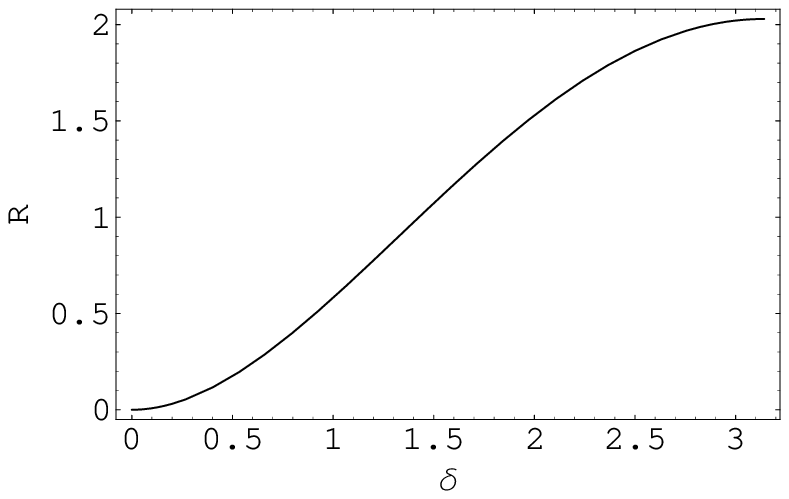,width=9cm}
\caption{\label{fin} Plot of $R$ as a function of $\delta$ in the range $[0,\pi]$.
 }}
The coefficient $c_0$ is easily determined and it is $1/(48)$.  The difference between $R$
and the polynomial $P$ is less than $10^{-8}$ over the whole range of the value of
$\delta$.

\noindent
Thus we have
\be
\begin{split}
W_2^{mnb}=&-
\frac{g^4(N^2-1)}{8\cdot 48\pi^4} \delta^2(2 \pi-\delta)^2=-\frac{g^4(N^2-1)}{384\pi^4} \delta^2(2 \pi-\delta)^2,
\end{split}
\ee
that exactly coincides with the second order expansion of the matrix model result (\ref{2d-result}), after the inclusion of the abelian contribution.

We have checked therefore the conjecture at two loops, at least for this particular class of loops. Now some remarks are in order: first of all we notice that in Feynman gauge double-exchange diagrams and the interactions conspire to reproduce the matrix model expression. This has to be contrasted with the circular Wilson loop case, where interactions vanish and simple exchange diagrams carry over the exact result. This suggests that probably there are more clever gauge choice to unveil the  birth of the matrix model, may be contour-dependent. The second observation is that our formulae allow easily to perform explicit computations for other contours: all the integrals are in fact finite, and it is just matter of patience to put them on computer and evaluate them numerically. We remark again that in spite of the presence of the cusp, these supersymmetric Wilson loops are indeed finite.

\section{Conclusions}
In this paper we have studied at quantum level a family of supersymmetric Wilson loops
in ${\cal N}=4$ SYM which have been proposed in \cite{Drukker:2007qr}. The contours
are restricted to an $S^2$ submanifold of space-time and then for a curve of
arbitrary shape a prescription for the scalar couplings guarantees that the resulting loop is globally
supersymmetric. We have shown that no divergence arises in the perturbative expansion, also when non-smooth "cusped"
curves are considered and we derived a general compact representation for the contribution of interactions at order $g^4$.
Finally we presented a two-loop computation, supporting the evidence that this class of loops, that in general
preserve four supercharges, may be described by a perturbative calculation in two-dimensional bosonic Yang-Mills theory on $S^2$. We chose
a ``wedge" contour and we performed the necessary analytical and numerical evaluation to find consistency with the conjecture proposed in
\cite{Drukker:2007qr}. The agreement with the prediction of the matrix model, describing the zero-instanton sector of Yang-Mills theory on $S^2$, is achieved in Feynman gauge after a non trivial conspiration of ladder and interacting diagrams: in this sense the situation here is quite different from the familiar circular Wilson loop, where exchange diagrams alone produce the correct matrix integral.

Many open questions remain of course to be answered: while our two-loop calculation points towards the validity of the correspondence between supersymmetric Wilson loops on $S^2$ and the gaussian matrix model of zero-instanton YM$_2$, further checks, involving other contours and, may be,
higher perturbative orders, are certainly welcome. More ambitiously one would like to generalize the localization approach on $S^4$, elaborated in \cite{Pestun:2007rz} for the circular loop, to the present situation, where a different amount of supersymmetry is preserved. In particular it would be important to understand the contributions of instantons in this case: instantons have been claimed to decouple in fact when the contour is a circle \cite{Pestun:2007rz}. Another direction of work is to study, in this context, correlators of Wilson loops, to understand if the matrix model/two-dimensional Yang Mills description still holds: in the same spirit it would be also interesting to study the correlators with chiral primary operators, as done in \cite{Semenoff:2006am} for a different class of loops. We end by observing that supersymmetry protects, in this case, cusped loops from being divergent: in a sense the BPS property smooths away the cusp anomalous behavior. It would be intriguing to understand better this last point and find, if any, connections among these exact results and the recent discovered perturbative properties of scattering amplitudes of ${\cal N}=4$ SYM. All these directions are currently under investigations.

\subsection*{Acknowledgments}
Two of us (DS and FP) would like thank to  Galileo Galilei Institute for
Theoretical Physics for hospitality  during the completion of this work.
DS and LG warmly thank N. Drukker for an illuminating discussion.

\newpage
\appendix
\addcontentsline{toc}{section}{Appendices}
\noindent{\Large\textbf{Appendices}}
\section{Light-cone analysis of planar Maldacena-Wilson loop}
\subsubsection*{ Contributions to the Wilson-loop at ${\cal O}(g^4)$ in light-cone gauge}
In this Appendix we present an analysis of the contributions at ${\cal O}(g^4)$
to a space-like planar  Maldacena-Wilson loop in light-cone gauge. We shall show
that the diagrams are separately finite and no divergence arises from the
integration over a smooth circuit.

\noindent
Let us briefly recall some basic definitions and notations.

\noindent
The light-cone gauge is characterized
by the introduction of a light-cone vector $n^{\mu},\,\, n^2=0$ leading to the
gauge condition $n^{\mu}A_{\mu}=0$, $A_{\mu}$ being the vector potential (internal
indices are understood). The free vector propagator in momentum space takes the form
$$D_{\mu \nu}(p)= \frac 1{p^2+i\epsilon} [g_{\mu \nu}-\frac {n_{\mu}p_{\nu}+
p_{\mu}n_{\nu}}{np+i\epsilon \hat{n}p}],$$
$\hat n$ being a null vector ($\hat{n}^2=0$) such that $n\hat{n}=1$. The usual choice
is $n^{\mu}=\frac 1{\sqrt 2}(1,0,0,-1)$ and $\hat{n}^{\mu}=\frac 1{\sqrt 2}(1,0,0,1)$.
The first term in the propagator corresponds to its expression in Feynman gauge,
whereas the second term is characteristic of the light-cone gauge.

\noindent
Since the loop we are concerned with lies on the plane orthogonal to the gauge vectors,
the ${\cal O}(g^4)$ contribution due to the exchange of two free propagators will
not be different from the one in Feynman gauge and thereby will not be
considered here.

\subsubsection*{\bf The self-energy correction.}
In ref.\cite{Bassetto:2007zj} the self-energy corrections to the gluon and scalar propagators
at ${\cal O}(g^2)$ for SUSY ${\cal N}=4$ have been computed. Since the Wilson
loop we are considering lies on the plane orthogonal to the gauge vectors $n$
and $\hat{n}$, it is easy to convince oneself that only ``transverse'' Green
functions contribute, which, in turn, get only $finite$ ${\cal O}(g^2)$
corrections in this gauge \cite{Mandelstam:1982cb}, at variance with the more familiar Feynman gauge
\cite{Erickson:2000af}. Since we expect that all the ${\cal O}(g^4)$ will be finite, there
will be no need of a dimensional regularization.
Typically in momentum space the gluon transverse Green function
receives the ${\cal O}(g^2)$ correction
\be
\label{trg}
G^{(2)}_{\alpha \beta}=-\delta_{\alpha \beta}\frac{ig^2N}{4\pi^2 p^2}[\frac{\pi^2}{6}-
L_2(1-\zeta)],
\ee
where $\alpha, \beta= 1,2$, $L_2$ is the Euler's dilogarithm and $\zeta = \frac{2np\,
\hat{n}p}{p^2}$.

\noindent
A direct calculation shows that the term containing the dilogarithm,  when Fourier
transformed and integrated over the contour, gives a vanishing contribution.
Then the one-loop correction to the single exchange is
\be
\begin{split}
S_2=
&\left(\frac{g^2}{8\pi^2}\right)^2(N^2-1)\left(\frac{\pi^2}{6}\right)
\oint\!\! d t_2d t_1
\frac{(\dot{x}_1\cdot \dot{x}_2)- |\dot{x}_1||\dot{x}_2|}{(x_1-x_2)^2}\\
\end{split}
\ee

\subsubsection*{\bf The triple vertex correction}
The triple vertex diagram appears in the expansion of the Wilson-loop at the order $g^4$. It
corresponds to compute (we have suppressed a total factor $-i g^3/N$, that will be inserted
back at the end of the calculation)
\be
\label{spider1a}
\begin{split}
S_3=&\int_0^1d{t}_1\int_0^{{t}_1} d{t}_2\int_0^{{t}_2} d{t}_3
\langle \mathrm{Tr}[\mathcal{A}({t}_1)\mathcal{A}({t}_2)\mathcal{A}({t}_3)]\rangle=\\
=&\frac{1}{3}\oint d{t}_1 d{t}_2 d{t}_3\eta({t}_1{t}_2{t}_3)
\mathrm{Tr}[\mathcal{A}({t}_1)\mathcal{A}({t}_2)\mathcal{A}({t}_3)]
 \end{split}
\ee
where
\be
\label{spider2a}
\mathcal{A}({t}_i)\equiv A_\mu(x({t}_i))x^\mu({t}_i)-i |\dot{x}({t}_i)|\Phi_\theta(x({t}_i)).
\ee
and $\eta(t_1,t_2,t_3)$ is defined in \eqref{spider3}.
Since the measure of integration $dt_1 dt_2 dt_3\eta(t_1,t_2,t_3)$ is invariant under cyclic permutations,
the integrand in  (\ref{spider1a}) can simply be written as
\be
\label{spider5a}
\begin{split}
\langle \mathrm{Tr}[\mathcal{A}(t_1)\mathcal{A}(t_2)\mathcal{A}(t_3)]\rangle=&
\dot{x}^\lambda(t_1)\dot{x}^\mu(t_2)
\dot{x}^\nu(t_3)
\langle \mathrm{Tr}[{A}_\lambda(t_1){A}_\mu(t_2){A}_\nu(t_3)]\rangle-\\
&-3\dot{x}^\lambda(t_1)|\dot{x}(t_2)||\dot{x}(t_3)|\langle
\mathrm{Tr}[{A}_\lambda(t_1)\Phi_\theta(t_2)\Phi_\theta(t_3)]\rangle.
\end{split}
\ee
To begin with, we  shall compute
\be
\begin{split}
&\langle\mathrm{Tr}[{A}_\lambda(t_1){A}_\mu(t_2){A}_\nu(t_3)]\rangle=
\mathrm{Tr}[T^a T^b T^c]\langle {A}^a_\lambda(t_1){A}^b_\mu(t_2){A}^c_\nu(t_3)\rangle
=\\
&=\mathrm{Tr}[T^a T^b T^c]\int \frac{d^{2\omega} p_1 d^{2\omega} p_2 d^{2\omega} p_3}{(2\pi)^{6\omega}}
e^{i p_1\cdot x_1+i p_2\cdot x_2+i p_3\cdot x_3}
\langle {A}^a_\lambda(p_1){A}^b_\mu(p_2){A}^c_\nu(p_3)\rangle_0.
\end{split}
\ee
We have introduced the short-handed notation $x^\mu_i\equiv x^\mu(t_i).$ The tree-level correlation function
$\langle {A}^a_\lambda(p_1){A}^b_\mu(p_2){A}^c_\nu(p_3)\rangle$ is given
in ref. \cite{Bassetto:2007zj}. We get
\be
\begin{split}
\langle {A}^a_\lambda(p_1)&{A}^b_\mu(p_2){A}^c_\nu(p_3)\rangle=
(2\pi)^{2\omega}\delta^{2\omega}(p_1+p_2+p_3)\frac{i g f^{abc}}{p_1^2 p_2^2 p_3^2}
\left[\delta_{\nu\lambda} \left(p_{3\mu} -\frac{p_{2\mu}}{p_{2}^+}p_3^+\right)+\right.\\
&+\mathrm{antisymmetrization\  in\   }(1\lambda)\ (2\mu)\ (3\nu) \Bigr],
\end{split}
\ee
with $p^+\equiv np$.
Then the contribution to the Wilson-loop is given by
\be
\label{3g}
\begin{split}
\frac{1}{3}&\oint dt_1 dt_2 dt_3\eta(t_1,t_2,t_3)\dot{x}_1^\lambda\dot{x}_2^\mu\dot{x}_3^\nu
\mathrm{Tr}[{A}_\lambda(t_1){A}_\mu(t_2){A}_\nu(t_3)]\rangle=\\
&=  i g f^{abc}\mathrm{Tr}[T^a T^b T^c]\oint dt_1 dt_2 dt_3\epsilon(t_1,t_2,t_3)
(\dot{x}_1\cdot\dot{x}_3)\times\\
&\times \dot{x}_2^\mu
\int
 \frac{d^{2\omega} p_2 d^{2\omega} p_3}{(2\pi)^{4\omega}}
\frac{e^{i p_2\cdot (x_2-x_1)+i p_3\cdot (x_3-x_1)}}{p_2^2 p_3^2(p_2+p_3)^2}
\left[ p_{3\mu} -
\frac{p_{2\mu}}{p_{2}^+}p_3^+\right],
\end{split}
\ee
where  $\epsilon(t_1,t_2,t_3)$ is the same function defined in sec. 3.
Eq.(\ref{3g}) can be then cast in the form
\be
\begin{split}
&\frac{1}{3}\oint dt_1 dt_2 dt_3\eta(t_1,t_2,t_3)\dot{x}_1^\lambda\dot{x}_2^\mu\dot{x}_3^\nu
\mathrm{Tr}[{A}_\lambda(t_1){A}_\mu(t_2){A}_\nu(t_3)]\rangle=\\
&=   g f^{abc}\mathrm{Tr}[T^a T^b T^c]\oint dt_1 dt_2 dt_3\epsilon(t_1,t_2,t_3)
(\dot{x}_1\cdot\dot{x}_3)\times\\
&\times\left(\dot{x}_2\frac{\partial \mathcal{I}_1(x_3-x_1,x_2-x_1)}{\partial x_3}-
\dot{x}_2\frac{\partial\mathcal{I}_2(x_3-x_1,x_2-x_1)}{\partial x_2}\right),
\end{split}
\ee
which implicitly defines the functions $\mathcal{I}_1(x_3-x_1,x_2-x_1)$ and
$\mathcal{I}_2(x_3-x_1,x_2-x_1)$. Actually for a planar space-like Maldacena-Wilson
loop the function $\mathcal{I}_2(x_3-x_1,x_2-x_1)$ can be rewritten in a way which is
manifestly Lorentz invariant. Any reference to the original light-like directions
disappears. The explicit expressions for $\mathcal{I}_1(x_3-x_1,x_2-x_1)$  and
$\mathcal{I}_2(x_3-x_1,x_2-x_1)$ are given in appendix B.

\noindent
Similarly, the scalar contribution turns out to be
\be
\begin{split}
&\oint d{t}_1 d{t}_2 d{t}_3\eta({t}_1{t}_2{t}_3)
\dot{x}^\lambda(t_1)|\dot{x}(t_2)||\dot{x}(t_3)|\langle
\mathrm{Tr}[{A}_\lambda(t_1)\Phi_\theta(t_2)\Phi_\theta(t_3)]\rangle=\\
&=g f^{abc}\mathrm{Tr}[T^a T^b T^c]
\oint d{t}_1 d{t}_2 d{t}_3\epsilon({t}_1{t}_2{t}_3)
|\dot{x}(t_1)||\dot{x}(t_3)|\times\\
&\times\left(\dot{x}_2\cdot\frac{\partial \mathcal{I}_1(x_3-x_1,x_2-x_1)}{\partial x_3}-
\dot{x}_2\cdot\frac{\partial \mathcal{I}_2(x_3-x_1,x_2-x_1)}{\partial x_2}\right),
 \end{split}
\ee
where symmetry properties of the integrals have been suitably taken into account.
Summing together vector and scalar contributions, we eventually obtain
\be
\label{v+s}
\begin{split}
S_3&= g f^{abc}\mathrm{Tr}[T^a T^b T^c]
\oint d{t}_1 d{t}_2 d{t}_3\epsilon({t}_1{t}_2{t}_3)
(\dot{x}_1\cdot\dot{ x}_3-|\dot{x}(t_1)||\dot{x}(t_3)|)\times\\
&\times\left(\dot{x}_2\cdot\frac{\partial \mathcal{I}_1(x_3-x_1,x_2-x_1)}{\partial x_3}-
\dot{x}_2\cdot\frac{\partial \mathcal{I}_2(x_3-x_1,x_2-x_1)}{\partial x_2}\right).
 \end{split}
\ee
This expression, a part from the kinematical factor, is identical to $\mathcal{A}$ in \eqref{spider5}.
From here on we can follow exactly the same steps taken in sec. 3. We shall again find that
$S_2$ cancels with a term coming from integrating by parts $S_3$.

\noindent
Eventually we are left with the gauge invariant result
\be
\label{finalformula}
\begin{split}
\mathcal{I}_{\mathrm{tot}} &=-\frac{ g^4 (N^2-1)}{128\pi^4}
\oint d{t}_1 d{t}_2 d{t}_3\epsilon({t}_1{t}_2{t}_3)
\frac{(\dot{x}_1\cdot\dot{ x}_3-|\dot{x}(t_1)||\dot{x}(t_3)|)}{(x_1-x_3)^2}\times\\
&\ \ \ \ \times \frac{(x_3-x_2)\cdot\dot{x}_2}{(x_3-x_2)^2}\log\left(\frac{(x_2-x_1)^2}{(x_3-x_1)^2}\right).
 \end{split}
\ee
We stress again that in light-cone gauge both self-energy correction and triple-vertex
contribution are {\it finite}, at variance with what happens in Feynman gauge;
there the two corrections exhibit a pole at $\omega=2$ and only in their sum the pole
cancels and the
{\it finite} quantity $\mathcal{I}_{\mathrm{tot}}$ is eventually recovered.

\noindent
It is almost trivial to realize that $\mathcal{I}_{\mathrm{tot}}$ vanishes identically when the
contour is a circle. Choosing the usual trigonometric parametrization for the circle, we can write
\be
\mathcal{I}_{\mathrm{tot}} =-\frac{ g^4(N^2-1)}{512\pi^4}
\oint d{t}_1 d{t}_2 d{t}_3\epsilon({t}_1{t}_2{t}_3)
 \cot \left(\frac{t_2-t_3}{2}\right) \log \left(\frac{1-\cos (t_1-t_2)}{1-\cos (t_3-t_1)}\right)=0.
\ee
since the integrand
\be
\epsilon({t}_1{t}_2{t}_3)\cot \left(\frac{t_2-t_3}{2}\right) \log \left(\frac{1-\cos (t_1-t_2)}{1-\cos (t_3-t_1)}\right)
\ee
is antisymmetric under the exchange $t_2\leftrightarrow t_3$, while the measure $d{t}_1 d{t}_2 d{t}_3$
and the region of integration are symmetric.

\section{Some properties of the integrals $\mathbf{\mathcal{I}_1(x,y)}$ and $\mathbf{\mathcal{I}_2(x,y)}$}
For the  integral  $\mathcal{I}_1(x,y)$  defined in \eqref{I1} one can easily perform the integration
over the momenta. In order to  integrate over $p_1$, we first introduce the Feynman
parametrization for the two denominators, which depends on $p_1$. Then we perform the
change of variable $p_1\mapsto p_1-\alpha p_2$. This yields
\be
\label{inizio}
\begin{split}
\mathcal{I}_1(x,y)&\equiv
\int \frac{d^{2\omega} p_1 d^{2\omega} p_2}{(2\pi)^{4\omega}}\frac{e^{i p_1 x+i p_2 y}}{p_1^2
p_2^2 (p_1+p_2)^2}
=\\
&=\int_0^1 d\alpha \int \frac{d^{2\omega} p_2 }{(2\pi)^{2\omega}}\frac{e^{i p_2 (y-\alpha x)}}{p_2^2}
\int \frac{d^{2\omega} p_1}{(2\pi)^{2\omega}}
\frac{e^{i p_1 x}}{[p_1^2+\alpha(1-\alpha)p_2^2]^2}
\end{split}
\ee
The integral over $p_1$ can be now evaluated by means of the Schwinger representation for the denominator
in \eqref{inizio}. We obtain
\be
\begin{split}
\mathcal{I}_1(x,y)&=\frac{1}{(4\pi)^\omega}\int_0^1 d\alpha  \int_0^\infty d\beta \beta^{1-\omega}
\int \frac{d^{2\omega} p_2 }{(2\pi)^{2\omega}}\frac{e^{i p_2 (y-\alpha x)}}{p_2^2}
{e^{-\frac{{x}^2}{4\beta}-\beta \alpha(1-\alpha)p_2^2  }}=\\
&=\frac{1}{(4\pi)^\omega}\int_0^1 d\alpha (\alpha(1-\alpha))^{\omega-2}  \int_0^\infty d\beta \beta^{1-\omega}
\int \frac{d^{2\omega} p_2 }{(2\pi)^{2\omega}}\frac{e^{i p_2 (y-\alpha x)}}{p_2^2}
{e^{-\frac{{x}^2}{4\beta}-\beta p_2^2  }}.
\end{split}
\ee
The integral over the second momentum can be now performed by introducing a second Schwinger parameter
$\lambda$. We end up with the following parametric representation for $\mathcal{I}_1(x,y)$
\be
\mathcal{I}_1(x,y)
 =\frac{ 1}{(4\pi)^{2\omega}}
\!\!\!\int_0^1\!\!\!\!
 d\alpha(\alpha(1-\alpha))^{\omega-2}\int_0^\infty d\beta
  \beta^{1-\omega}\int_0^\infty d\lambda (\lambda+\beta)^{-\omega}
  e^{-\frac{({y}-\alpha {x})^2}{4(\lambda+\beta)}-\frac{{x}^2\alpha(1-\alpha)}{4\beta}}.
\ee
By setting $\tau=\lambda+\beta$, we can first integrate over $\beta$ and then over $\tau$. In fact
\be
\begin{split}
\mathcal{I}_1(x,y)
  &=\frac{ 1}{(4\pi)^{2\omega}}
\!\!\!\int_0^1\!\!\!\!
 d\alpha(\alpha(1-\alpha))^{\omega-2}\int_0^\infty d\tau \tau^{-\omega}
 \int_0^\tau d\beta
  \beta^{1-\omega}
  e^{-\frac{({y}-\alpha {x})^2}{4\tau}-\frac{{x}^2\alpha(1-\alpha)}{4\beta}}=\\
&=\frac{ 4^{\omega -2} ({x}^2)^{2-\omega } }{(4\pi)^{2\omega}}
\!\!\!\int_0^1\!\!\!\!
 d\alpha\int_0^\infty d\tau \tau^{\omega-2}  e^{-\frac{({y}-\alpha {x})^2}{4}\tau}
  \Gamma \left(\omega
   -2,\frac{{x}^2 (1-\alpha ) \alpha\tau }{4 }\right)=\\
&=\frac{\Gamma(2\omega-3)}{64\pi^{2\omega}(\omega-1)}
\int_0^1
 d\alpha~~
 \frac{[\alpha(1-\alpha)]^{\omega-2}}{\left[\alpha (x-y)^2+(1-\alpha) y^2\right]^{2\omega-3}}\times\\
&\ \ \ \ \ \ \ \ \times{_2F_1}
\left(1,2\omega-3,\omega,\frac{({y}-\alpha{x})^2}{\alpha (x-y)^2+(1-\alpha) y^2}\right).
 \end{split}
\ee
In the last equality, we have used the following integral given on the table
\be
\int^\infty_0 x^{\mu-1} e^{-\beta x} \Gamma(\nu,\alpha x) dx=
\frac{\alpha^\nu\Gamma(\mu+\nu)}{\mu(\alpha+\beta)^{\mu+\nu}} {_2F_1}
\left(1,\mu+\nu,\mu+1,\frac{\beta}{\beta+\alpha}\right).
\ee
This representation is useful to study the behavior around $x=0$, $y=0$ and
$y=x$. Since \eqref{inizio} is manifestly symmetric in the exchange $x\leftrightarrow y$
and $x\leftrightarrow y-x$, it is sufficient to study the behavior only around
$x=0$. The other two cases will obviously display a similar behavior. At $x=0$ we find
\be
\begin{split}
\mathcal{I}_1(0,y)&=\frac{\Gamma(2\omega-3){_2F_1}
\left(1,2\omega-3,\omega,1\right)}{64\pi^{2\omega}(\omega-1)\left[
 {y}^2\right]^{2\omega-3}}
\!\!\!\int_0^1\!\!\!\!
 d\alpha {[\alpha(1-\alpha)]^{\omega-2}}=\\
 &=\frac{ \Gamma^2 (\omega -1)}{64 \pi ^{2 \omega }
 (2 \omega-3)(2-\omega)}
  \frac{1}{\left[
 {y}^2\right]^{2\omega-3}}
 \end{split}
 \ee
The integral  $\mathcal{I}_2(x,y)$ is defined as follows
\be
\label{pipp}
\begin{split}
 \mathcal{I}_2(x,y)
 &=-\frac{ \Gamma(2\omega-3) }{64\pi^{2\omega}(\omega-1)}
\!\!\!\int_0^1\!\!\!\!
 d\alpha \frac{\alpha^{\omega-1}(1-\alpha)^{\omega-2}}{\left[\alpha(1-\alpha){x}^2+
 ({y}-\alpha{x})^2\right]^{2\omega-3}}\times\\
&\ \ \ \ \ \ \ \ \ \ \ \ \ \ \ \ \ \ \ \ \ \ \ \ \ \ \ \ \ \times{_2F_1}
\left(1,2\omega-3,\omega,\frac{({y}-\alpha{x})^2}{({y}-\alpha{x})^2+
\alpha(1-\alpha){x}^2}\right).
 \end{split}
\ee
The origin of this object is explained in appendix A and it is related to light-cone
gauge analysis of the Wilson-loop. There, its definition is given in momentum space.
The expression \eqref{pipp} is obtained performing the integration over the momenta along
the same path followed for $\mathcal{I}_1(x,y)$.

\noindent
In the following  we shall compute  its  behavior  at $x=0$, $y=0$ and $y=x$.
\noindent
At ${x}=0$:
\be
\begin{split}
 \mathcal{I}_2(x,y)
 &=-\frac{\Gamma(2\omega-3){_2F_1}
\left(1,2\omega-3,\omega,1\right)}{64\pi^{2\omega}\left[
 (\tilde{y})^2\right]^{2\omega-3}(\omega-1)}
\!\!\!\int_0^1\!\!\!\!
 d\alpha {\alpha^{\omega-1}(1-\alpha)^{\omega-2}}{}=\\
 &=-\frac{\Gamma^2(\omega-1)}{128\pi^{2\omega}(2-\omega)(2\omega-3)\left[
 ({y})^2\right]^{2\omega-3}}
=\\
&=\frac{1}{128 \pi ^4 (\omega -2){y}^2}+O\left((\omega
   -2)^0\right)
 \end{split}
\ee
At $y=0$:
\be
\begin{split}
\mathcal{I}_2(x,y)&=-\frac{ \Gamma(2\omega-3) }{64\pi^{2\omega}(\omega-1)\left[x\right]^{2\omega-3}}
\!\!\!\int_0^1\!\!\!\!
 d\alpha {\alpha^{2-\omega}(1-\alpha)^{\omega-2}}{_2F_1}
\left(1,2\omega-3,\omega,\alpha\right)=\\
&=-\frac{ \Gamma(2\omega-3) \Gamma(3-\omega)\Gamma(\omega-1)}{64\pi^{2\omega}(\omega-1)\left[{x}\right]^{2\omega-3}}
{_3F_2}(1,2\omega-3,3-\omega;\omega,2|1)=\\
&=-\frac{ \Gamma(2\omega-3) \Gamma(3-\omega)\Gamma(\omega-1)}{64\pi^{2\omega}\left[{x}\right]^{2\omega-3}}
\frac{ (\Gamma (\omega -2)-2  \Gamma (3-\omega ) \Gamma (2 \omega -4))}{4 (\omega -2)^3 \Gamma (2-\omega ) \Gamma (2 \omega
   -4)}=\\
   &=-\frac{1}{384 \pi ^2 {x}^2}+O\left((\omega -2)^1\right)
 \end{split}
\ee
At $y=x$:
\be
\begin{split}
\mathcal{I}_2
 &=-\frac{ \Gamma(2\omega-3) }{64\pi^{2\omega}(\omega-1)[{x}^2]^{2\omega-3}}
\!\!\!\int_0^1\!\!\!\!
 d\alpha {\alpha^{\omega-1}(1-\alpha)^{1-\omega}}\times{_2F_1}
\left(1,2\omega-3,\omega,1-\alpha\right)=\\
&=-\frac{ \Gamma(2\omega-3)\Gamma(2-\omega)\Gamma(\omega) }{64\pi^{2\omega}(\omega-1)[{x}^2]^{2\omega-3}}
{_3F_2}(1,2\omega-3,2-\omega;\omega,2|1)=\\
&=-\frac{ \Gamma(2\omega-3)\Gamma(2-\omega)\Gamma(\omega) }{64\pi^{2\omega}(\omega-1)[{x}^2]^{2\omega-3}}
\frac{1-\frac{\Gamma (\omega -1)}{\Gamma (3-\omega ) \Gamma (2 \omega -2)}}{2 (\omega -2)}=\\
   &=\frac{1}{64 \pi ^4 (\omega -2){x}^2}+O\left((\omega
   -2)^0\right)
 \end{split}
\ee
\paragraph{A useful combination of $\mathcal{I}_1$ and $\mathcal{I}_2$:}
In the following we shall show that the following combination of the derivatives of
$\mathcal{I}_1$ and $\mathcal{I}_2$,
\be
V_\mu=\frac{\partial \mathcal{I}_1({x},{y})}{\partial {x}^\mu}-
\frac{\partial \mathcal{I}_2({x},{y})}{\partial {y}^\mu},
\ee
can be reduced to a very simple form. First, we shall take the derivative.
We find
\be
\begin{split}
V_\mu&=\frac{\Gamma(2\omega-3)}{64\pi^{{2\omega}}(\omega-1)}
\!\!\!\int_0^1\!\!\!\!
 d\alpha
\left(\frac{\partial}{\partial {x}^\mu}+\alpha
\frac{\partial}{\partial {y}^\mu}\right)\left[
\frac{[\alpha(1-\alpha)]^{\omega-2}}{\left[(\tilde{y}-\alpha\tilde{x})^2\right]^{2\omega-3}}
G[\xi]
\right]=\\
&=\frac{\Gamma(2\omega-3)}{64\pi^{{2\omega}}(\omega-1)}
\!\!\!\int_0^1\!\!\!\!
 d\alpha
\frac{[\alpha(1-\alpha)]^{\omega-2}}{\left[(\tilde{y}-\alpha\tilde{x})^2\right]^{2\omega-3}}
G^\prime[\xi]\left(\frac{\partial\xi}{\partial {x}^\mu}+\alpha
\frac{\partial\xi}{\partial {y}^\mu}\right),
\end{split}
\ee
 where
\be
\xi=\frac{(\tilde{y}-\alpha{x})^2}{\alpha(1-\alpha){x}^2+
 (\tilde{y}-\alpha{x})^2}\ \ \ \ \mathrm{and}\ \ \ \ \
G[\xi]=\xi^{2\omega-3}{_2F_1}
\left(1,2\omega-3,\omega,\xi\right).
\ee
Since
\be
\left(\frac{\partial\xi}{\partial {x}^\mu}+\alpha
\frac{\partial\xi}{\partial {y}^\mu}\right)=
-2 (1-\xi) \xi \frac{{x}^\mu}{{x}^2},
\ee
the expression for $V_\mu$ can be rewritten as follows
\be
\begin{split}
V_\mu
&=-\frac{2\Gamma(2\omega-3)\tilde{x}^\mu}{64\pi^{{2\omega}}(\omega-1) (\tilde{x}^2)^{2\omega-2}}
\!\!\!\int_0^1\!\!\!\!
 d\alpha
{[\alpha(1-\alpha)]^{1-\omega}}{}
G^\prime[\xi]\xi^{4-2\omega} (1-\xi)^{2\omega-2}.
\end{split}
\ee
The derivative of $G[\xi]$ can be now computed by using the well-known properties of the
hypergeometric functions:
\be
\begin{split}
G^\prime(\xi)
&=\frac{2\omega-3}{\omega}\xi^{2\omega-4}(
\omega {_2F_1}
\left(1,2\omega-3,\omega,\xi\right)+\xi { \, _2F_1(2,2 \omega -2;\omega +1;\xi )} )=\\
&=(2\omega-3)\xi^{2\omega-4}{_2F_1}(1,2\omega-2;\omega;\xi )
\end{split}
\ee
where we have used  the identity
\be
\gamma {_2F_1}(\alpha,\beta;\gamma;\xi )-\gamma {_2F_1}(\alpha,\beta+1;\gamma;\xi )
+\alpha \xi {_2F_1}(\alpha+1,\beta+1;\gamma+1;\xi )=0.
\ee
Thus
\be
\begin{split}
V_\mu
&=-\frac{\Gamma(2\omega-2){x}^\mu}{32\pi^{{2\omega}}(\omega-1) ({x}^2)^{2\omega-2}}
\!\!\!\int_0^1\!\!\!\!
 d\alpha
{[\alpha(1-\alpha)]^{1-\omega}}{}
{_2F_1}(1,2\omega-2;\omega;\xi ) (1-\xi)^{2\omega-2}.
\end{split}
\ee

\end{document}